\documentclass[12pt]{article}
\usepackage{jheppub}
\usepackage{mathtools,amssymb,amsthm}
\usepackage[utf8]{inputenc}
\usepackage{enumitem}
\usepackage{amsmath}
\usepackage{amssymb}
\usepackage[all]{xy}
\usepackage{xcolor}
\usepackage{tikz}
\usepackage{microtype}
\usepackage{ragged2e}

\usepackage{babel}
\begin{document}
\global\long\def\aad{(a\tilde{a}+a^{\dagger}\tilde{a}^{\dagger})}%

\newcommand{\su}{\mathfrak{su}(1,1)}
\newcommand{\mbc}{\mathbb{C}}
\newcommand{\ads}{\text{AdS}}
\newcommand{\mh}{\mathcal{H}}
\newcommand{\mbp}{\mathbb{P}}
\newcommand{\me}{\mathcal{E}}
\newcommand{\uqsu}{\mathcal{U}_q (\mathfrak{su}(1,1))}
\newcommand{\uqsl}{\mathcal{U}_q (\mathfrak{sl}_2)}
\newcommand{\uqslr}{\mathcal{U}_q (\mathfrak{sl}(2,\mathbb{R}))}
\newcommand{\dplus}{D^{+}_\Delta}
\newcommand{\Vd}{\hat{V}_\Delta}

\global\long\def\ad{{\rm ad}}%

\global\long\def\bij{\langle ij\ket}%

\global\long\def\df{\coloneqq}%

\def\JX#1{{\color{purple}{[#1]}}}
\def\jx#1{{\color{purple}{[#1]}}}

\def\jj#1{{\color{blue}{[#1]}}}

\global\long\def\bra{\langle}%

\global\long\def\dd{{\rm d}}%

\global\long\def\dg{{\rm {\rm \dot{\gamma}}}}%

\global\long\def\ddt{\frac{{\rm d^{2}}}{{\rm d}t^{2}}}%

\global\long\def\ddg{\nabla_{\dot{\gamma}}}%

\global\long\def\del{\mathcal{\delta}}%

\global\long\def\Del{\Delta}%

\global\long\def\dtau{\frac{\dd^{2}}{\dd\tau^{2}}}%

\global\long\def\ul{U(\Lambda)}%

\global\long\def\udl{U^{\dagger}(\Lambda)}%

\global\long\def\dl{D(\Lambda)}%

\global\long\def\da{\dagger}%

\global\long\def\id{{\rm id}}%

\global\long\def\ml{\mathcal{L}}%

\global\long\def\mm{\mathcal{\me}}%

\global\long\def\mf{\mathcal{\mathcal{F}}}%

\global\long\def\ket{\rangle}%

\global\long\def\kpp{k^{\prime}}%

\global\long\def\lr{\leftrightarrow}%

\global\long\def\lf{\leftrightarrow}%

\global\long\def\ma{\mathcal{A}}%

\global\long\def\mb{\mathcal{B}}%

\global\long\def\md{\mathcal{D}}%

\global\long\def\mbr{\mathbb{R}}%

\global\long\def\mbz{\mathbb{Z}}%

\global\long\def\mh{\mathcal{\mathcal{H}}}%

\global\long\def\mi{\mathcal{\mathcal{I}}}%

\global\long\def\ms{\mathcal{\mathcal{\mathcal{S}}}}%

\global\long\def\mg{\mathcal{\mathcal{G}}}%

\global\long\def\mfa{\mathcal{\mathfrak{a}}}%

\global\long\def\mfb{\mathcal{\mathfrak{b}}}%

\global\long\def\mfb{\mathcal{\mathfrak{b}}}%

\global\long\def\mfg{\mathcal{\mathfrak{g}}}%

\global\long\def\mj{\mathcal{\mathcal{J}}}%

\global\long\def\mk{\mathcal{K}}%

\global\long\def\mmp{\mathcal{\mathcal{P}}}%

\global\long\def\mn{\mathcal{\mathcal{\mathcal{N}}}}%

\global\long\def\mq{\mathcal{\mathcal{Q}}}%

\global\long\def\mo{\mathcal{O}}%

\global\long\def\qq{\mathcal{\mathcal{\mathcal{\quad}}}}%

\global\long\def\ww{\wedge}%

\global\long\def\ka{\kappa}%

\global\long\def\nn{\nabla}%

\global\long\def\nb{\overline{\nabla}}%

\global\long\def\pathint{\langle x_{f},t_{f}|x_{i},t_{i}\ket}%

\global\long\def\ppp{p^{\prime}}%

\global\long\def\qpp{q^{\prime}}%

\global\long\def\we{\wedge}%

\global\long\def\pp{\prime}%

\global\long\def\sq{\square}%

\global\long\def\vp{\varphi}%

\global\long\def\ti{\widetilde{}}%

\global\long\def\wg{\widetilde{g}}%

\global\long\def\te{\theta}%

\global\long\def\tr{{\rm Tr}}%

\global\long\def\ta{{\rm \widetilde{\alpha}}}%

\global\long\def\sh{{\rm {\rm sh}}}%

\global\long\def\ch{{\rm ch}}%

\global\long\def\Si{{\rm {\rm \Sigma}}}%

\global\long\def\si{{\rm {\rm \sigma}}}%

\global\long\def\sch{{\rm {\rm Sch}}}%

\global\long\def\vol{{\rm {\rm {\rm Vol}}}}%

\global\long\def\reg{{\rm {\rm reg}}}%

\global\long\def\zb{{\rm {\rm |0(\beta)\ket}}}%

\newcommand{\be}{\begin{equation}}
\newcommand{\ee}{\end{equation}}
\newcommand{\tx}{\text}
\newcommand{\tb}{\textbf}
\title{Quantum Symmetry and Geometry in Double-Scaled SYK}
\author[a]{Jeremy van der Heijden,}

\author[b]{Erik Verlinde,}

\author[c]{Jiuci Xu}

\affiliation[a]{Department of Physics and Astronomy, University of British Columbia, 6224 Agricultural Road, Vancouver, B.C. V6T 1Z1, Canada}
\emailAdd{jeremy.vanderheijden@ubc.ca}

\affiliation[b]{Institute for Theoretical Physics, University of Amsterdam, Science Park 904, Postbus 94485, 1090 GL Amsterdam, The Netherlands}
\emailAdd{E.P.Verlinde@uva.nl}

\affiliation[c]{Department of Physics, University of California, Santa Barbara, CA 93106, USA}
\emailAdd{Jiuci\_Xu@ucsb.edu}

\abstract{The emergence of the quantum $R$-matrix in the double-scaled SYK model points to an underlying quantum group structure. In this work, we identify the quantum group $\mathcal{U}_q(\mathfrak{su}(1,1))$ as a subalgebra of the chord algebra. Specifically, we construct the generators of $\uqsu$ from combinations of operators within the chord algebra and show that the one-particle chord Hilbert space decomposes into the positive discrete series representations of $\mathcal{U}_q(\mathfrak{su}(1,1))$. Using the coproduct structure of the quantum group, we build the multi-particle Hilbert space and establish its equivalence with previous results defined by the chord rules. In particular, we show that the quantum $R$-matrix acts as a swapping operator that reverses the ordering of open chords in each fusion channel while incorporating the corresponding $q$-weighted penalty factors. This action enables an explicit derivation of the chord Yang–Baxter relation. We further explore a realization of the quantum group generators on the quantum disk, and present a novel factorization formula for the bulk gravitational wavefunction in the presence of matter. We further discuss the relation between the $\uqsu$ structure uncovered here and the $\uqslr$ algebra previously studied from the boundary perspective. Finally, we study the gravitational wavefunction with matter in the Schwarzian regime.

}

\maketitle

\section{Introduction}

In quantum field theory and gravity, symmetry constrains possible interactions and organizes the spectrum of excitations into representations of the underlying symmetry algebra.  
Within the framework of the AdS/CFT correspondence~\cite{Maldacena:1997re,Susskind:1994vu,Witten:1998zw,Susskind:1998dq,Gubser_1998}, symmetry plays an even deeper role: it provides a guiding principle for how bulk gravitational physics is encoded in the dual boundary description~\cite{Harlow:2018jwu,Harlow:2018tng,Benini_2023}.  

Recently, the double-scaled SYK (DSSYK) model~\cite{Cotler_2017,Berkooz:2018qkz,Berkooz:2018jqr,Berkooz:2024lgq} has attracted significant attention as a solvable system with a proposed gravitational dual that extends from the semiclassical to the fully quantum regime~\cite{susskind:2023hnj,susskind:2023rxm,Okuyama:2023bch,narovlansky:2023lfz,verlinde:2024znh,Blommaert:2023opb,Blommaert:2023wad,Blommaert:2024whf,Aguilar-Gutierrez:2024nau,Aguilar-Gutierrez:2024oea,milekhin:2023bjv,Almheiri:2024ayc,Sekino:2025bsc}. 
However, a unified understanding of this duality remains incomplete, partly due to the rich algebraic and dynamical structure of the model~\cite{Lin_2023,xu2024von}.   In particular, incorporating matter chords reveals intricate interactions and symmetries that are difficult to reproduce consistently on the gravitational side across the full parameter space of the theory~\cite{Blommaert:2024whf,Blommaert:2025avl}. Tremendous progress has also been made in understanding the dual gravitational description of DSSYK in the semiclassical regime~\cite{Rabinovici:2023yex,Ambrosini:2024sre,Rahman:2024vyg,Rahman:2024iiu,Milekhin:2024vbb,Aguilar-Gutierrez:2025hty,okuyama2023end,Okuyama:2025hsd,Heller:2024ldz,Aguilar-Gutierrez:2025pqp,Narovlansky:2025tpb,Miyaji:2025yvm,Xu:2025chord,Ambrosini:2025hvo}. For recent progress in incorporating matter fields into the gravitational description, yielding results consistent with DSSYK in the quantum regime, see~\cite{Cui:2025sgy}.

Motivation for identifying a consistent bulk dual arises from the existence of an intrinsic notion of a “bulk” Hilbert space within DSSYK itself.  
This idea builds on the pioneering work of~\cite{Berkooz:2018jqr,Berkooz:2018qkz}, which introduced chord diagrams as a tool to make the model’s solvability manifest.  
In this formulation, observables are represented by operators placed on a circle, and their correlation functions are computed by summing over all Wick contractions, equivalently over all possible chord configurations connecting identical type of operators.  
Each crossing of two chords is weighted by a factor determined by the ratio of number of fermions contained in each operator in the original SYK model. The weights, together with the types of chord operators, effectively define the model itself, independent of its microscopic origin in the original SYK framework.

The chord picture naturally introduces an inner-product structure on the space of states accounting for in-equivalent open chord configurations\footnote{We review this notion as well as the construction of the Hilbert space in section \ref{sec:review}.}, giving rise to an intrinsic notion of a bulk Hilbert space.  
Analogous to the Euclidean path integral in semiclassical gravity, one can “slice” a chord diagram to define a bulk state: open chords intersecting the slice define the state’s boundary data, while summing over all crossings yields their inner products.  
Remarkably, this bulk description is intrinsically discrete: configurations with a fixed number of open chords form a finite-dimensional subspace for a given set of chord operators.  In the triple-scaled limit~\cite{Lin:2022rbf}, where the chord number is taken to infinity, the discrete bulk Hilbert space reduces to the continuum Hilbert space describing the renormalized geodesic length between two boundaries in empty AdS$_2$. This correspondence can be further extended to include the case with $\mathfrak{sl}_2$ matter excitations~\cite{Harlow:2021dfp,Xu:2024gfm}.    
This discrete bulk structure is one of the model’s distinctive features, and various gravitational realizations have been proposed to capture it, including sine-dilaton gravity~\cite{Blommaert:2023opb,Blommaert:2024ymv,Mertens:2025qg1}, the holographic quantum disk~\cite{Almheiri:2024ayc}, and more recently, complex Liouville string theory~\cite{Blommaert:2025eps}.  

Despite the challenges of realizing DSSYK’s full quantum regime in a gravitational description, evidence suggests that an underlying quantum group organizes these structures and accounts for the intrinsic discreteness of the bulk Hilbert space~\cite{Berkooz:2022mfk,Lin_2023,Mertens:2025qg1,Mertens:2025qg2}.  
It is therefore essential to clarify how this symmetry emerges from the boundary formulation.  
The symmetry structure of DSSYK has previously been analyzed in terms of chord operators in~\cite{Lin_2023}, where a $U(J)$ algebra, interpreted as a $q$-analogue of the universal $\mathfrak{sl}_2$ algebra,\footnote{Here we refer to the physical $\mathfrak{sl}_2$ algebra generated by diffeomorphism-invariant bulk operators, not the boundary gauge symmetry $SL(2,\mathbb{R})$. For clarification, see \cite{Lin:2019qwu}.} was shown to capture the sub-maximal Lyapunov exponent in the semiclassical limit at finite temperature.\footnote{This regime includes and extends the aforementioned triple-scaled limit beyond the low-temperature approximation.}  
However, it remains unclear how the quantum $R$-matrix of $\uqsu$ arises from crossed matter-chord configurations with fixed boundary energy~\cite{askey1994kernel}, and how the quantum group structure more generally underlies the chord description and its discrete bulk features.   Addressing these questions forms the main motivation of the present work.

\paragraph{Goal of the Paper}

In this work, we investigate the quantum group structure underlying the chord Hilbert space of the double-scaled SYK model with matter.  
Within the one-particle chord Hilbert space, we identify specific combinations of chord ladder operators and their associated number operators that reproduce the generators of $\mathcal{U}_q(\mathfrak{su}(1,1))$.  
We further demonstrate that the resulting positive structure is compatible with the chord inner product.  
Building on this observation, we show that the one-particle Hilbert space decomposes into a tensor product of positive discrete series representations of the quantum group, each closed under the action of its generators.  

We then extend this structure to the multi-particle sector by utilizing the coproduct of the quantum group, which can be naturally realized as a coproduct within the two-sided chord algebra generated by the left and right 
$q$-ladder operators.  Using this framework, we construct multi-particle states and present an explicit unitary intertwiner that decomposes a two-particle state into the fusion of two one-particle states.  
We also examine the action of the quantum $R$-matrix of $\uqsu$ on these states, showing that it reverses the ordering of chords in the intermediate channel and keeps track of the penalty factors associated with the  crossings generated correspondingly.  

As an application, we demonstrate that the chord Yang–Baxter equation, previously regarded as an obstruction in identifying a chord as an intrinsic geometrical object~\cite{Lin_2023}, naturally follows from the fusion rules of the quantum group. 
In particular, the pentagon identity for the quantum $6j$-symbol \cite{jafferis2023} can be interpreted as a diagrammatic identity equating a loop-level chord configuration in the bulk with a tree-level diagram, where all intermediate energies are placed ``on shell'' by the boundary conditions.  
From this viewpoint, bulk interactions are encoded in the fusion rules of the quantum group, while the Yang–Baxter relation emerges as a consequence of their consistency conditions and unitarity.  

Finally,  We provide a novel factorization property of the gravitational wavefunction in the quantum regime, in the spirit of~\cite{Mertens:2025qg1}, which raises intriguing questions about the interplay between two distinct quantum groups, $\uqsu$ and $\uqslr$, in the DSSYK model.
We then analyze the semiclassical limit of the quantum group structure and provide new expressions for the limiting gravitational wavefunctions coupled to matter.  

\paragraph{Organization of the Paper}

In \textbf{Section}~\ref{sec:review}, we review the chord description of the double-scaled SYK model.  
In particular, we discuss the construction of the zero- and one-particle chord Hilbert spaces and introduce the notion of the chord intertwiner, which identifies the one-particle Hilbert space as the space of operators acting on the zero-particle sector.  
This perspective facilitates the later construction of quantum group generators in terms of chord ladder operators.  

In \textbf{Section}~\ref{sec:qg-structure}, we construct the generators of $\mathcal{U}_q(\mathfrak{su}(1,1))$ explicitly in terms of chord ladder operators and show that they realize the positive discrete-series representation $\mathcal{D}^{+}_\Delta$ of $\uqsu$ on a specific set of states in the one-particle Hilbert space.  
This leads to a decomposition of $\mathcal{H}_1$ into a tensor product of $\mathcal{D}^{+}_\Delta$ representations, each weighted by the energy density of the pure DSSYK model\footnote{By pure DSSYK we mean the model in absence of matter, and the energy density is defined by the overlap of energy eigenstates. }.  
We further demonstrate that the $*$-structure of the quantum group agrees with the Hermitian conjugation of chord operators, and that the co-product structure translates into one acting on the two-sided algebra generated by left and right chord ladders.  
Using this structure, we construct the multi-particle Hilbert space; in the two-particle sector, we present an explicit unitary intertwiner that decomposes a two-particle state into the fusion of two one-particle states.  
We then show that the action of the quantum $R$-matrix reverses the ordering of chords in the intermediate channel and incorporates the associated penalty factors for crossings.  
The two-particle Hilbert space obtained from quantum group representations coincides with that defined by chord combinatorics, and we further extend this correspondence to the full multi-particle Hilbert space.

\justifying

In \textbf{Section}~\ref{sec:semi-classical}, we examine the bulk interpretation of the gravitational wavefunctions coupled to matter constructed in the previous section.  
We develop a $q$-Jackson integral representation and derive a novel factorization formula that expresses the bulk wavefunction as a $q$-convolution of two ``boundary wavefunctions,'' following the spirit of~\cite{Mertens:2025qg1}. This observation motivates a speculative discussion on the possible relation between the $\uqsu$ structure uncovered in this work and the $\uqslr$ algebra identified from the boundary perspective in~\cite{Mertens:2025qg1}.  We emphasize that we do not claim a direct correspondence between the two. Rather, we speculate that their distinction may originate from the two $\mathfrak{sl}_2$ algebras appearing in JT gravity: the gauge $\mathfrak{sl}_2$ algebra associated with the extended Hilbert space consisting of boundary Schwarzian particles and bulk matter, and the  physical $\mathfrak{sl}_2$ algebra constructed in~\cite{Lin:2019qwu}, which acts on the physical bulk Hilbert space. In the double-scaled SYK model, these algebras may be promoted to $\uqsl$ under $q$-deformation, but acquire different positive structures, giving rise respectively to the $\uqslr$ algebra discussed in~\cite{Mertens:2025qg1} and the $\uqsu$ algebra studied in the present paper. These two quantum groups may thus play distinct and complementary roles in the full description of DSSYK, although clarifying this connection remains an open question.

Finally, in \textbf{Section}~\ref{sec:discussion}, we summarize our results and discuss several directions for future investigation. 
Various technical details and intermediate derivations are collected in the appendices.

\section{Brief Review of Chord Description of DSSYK} \label{sec:review}

In this section, we review the chord description of the double-scaled SYK model and its bulk Hilbert space interpretation.
We begin with the case involving a single type of chord operator, referred to as the $H$-chord, and construct the corresponding zero-particle Hilbert space $\mathcal{H}_0$.  
We then extend the framework by introducing an additional type of chord operator, the $M$-chord, which plays the role of matter insertion, analogous to coupling JT gravity to matter.  
The resulting chord algebra naturally leads to a bulk Hilbert space with a single matter insertion, denoted $\mathcal{H}_1$.  
We introduce a useful notion of an intertwiner that factorizes $\mathcal{H}_1$ isometrically into two copies of $\mathcal{H}_0$, i.e.~$\mathcal{H}_1 \simeq \mathcal{H}_0 \otimes \mathcal{H}_0$, a construction that will facilitate our later analysis of the quantum group structure.  
Finally, we identify a particular class of states in $\mathcal{H}_1$ that satisfy a remarkable recursion relation, signaling the emergence of an underlying quantum group symmetry.

\subsection{The Chord Description of the Double-Scaled SYK Model}

In this section, we review the double-scaled SYK model, focusing on the formulation in terms of a single type of chord operator $H$, referred to as the Hamiltonian chord operator.  
This operator arises as the double-scaling limit of the original SYK Hamiltonian $H_{\text{SYK}}$, such that the expectation value of $H^{k}$ in the ``empty state'' $|0\rangle$—defined later in this section—reproduces the disorder-averaged moments of the same order of the SYK Hamiltonian in this limit:  
\be \label{eq:doublescaling}
\overline{\text{Tr}\!\left(H_{\text{SYK}}^{k}\right)} 
\;\stackrel{\text{Double-Scaling Limit}}{\Longrightarrow}\;
\langle 0| H^{k} |0\rangle~, 
\quad k = 0, 1, 2, \dots ,
\ee
where the overline denotes disorder averaging, and $\text{Tr}$ is the trace in the original SYK model.  
For modern reviews, see~\cite{Lin:2022rbf,Berkooz:2024lgq}.  
In the following, we will not need the detailed definition of the double-scaling limit and will focus on its chord description and the associated Hilbert space.  

The expectation value of the $k$-th moment of $H$ in the empty state can be computed using chord rules \cite{Berkooz:2018jqr}:  
one places $k$ marked points on the boundary of a circle and sums over all possible Wick contractions, which are represented by lines (chords) connecting identical operator insertions through the interior of the circle.  
Each pair of chords is allowed to cross at most once, and each crossing contributes a factor of $q^2$, where $0\leq q<1$\footnote{The choice of $q^{2}$, rather than the more conventionally used $q$ in DSSYK literature, is made to align with the conventions used later in the representation theory of the quantum group $\mathcal{U}_{q}(\mathfrak{su}(1,1))$.}.  
Thus, a given contraction corresponds to a unique chord diagram, contributing a weight $q^{2\text{cr}}$, where $\text{cr}$ denotes the number of crossings.  
The case $k=0$ corresponds to the empty diagram, which contributes $\bra 0|0\ket = 1$.  

As an example, the fourth moment of $H$ is given by
\be 
\langle 0| H^{4} | 0\rangle =
\begin{tikzpicture}[baseline={([yshift=-0.1cm]current bounding box.center)},scale=.7]
    \draw[black] (0,0) circle (1);
    \draw (-1,0) .. controls (0,0) .. (0,1);
    \draw (1,0) .. controls (0,0) .. (0,-1);
    \node at (-1,0) [circle,fill,black,inner sep=1.pt]{};
    \node at (1,0) [circle,fill,black,inner sep=1.pt]{};
    \node at (0,1) [circle,fill,black,inner sep=1.pt]{};
    \node at (0,-1) [circle,fill,black,inner sep=1.pt]{};
\end{tikzpicture}
+
\begin{tikzpicture}[baseline={([yshift=-0.1cm]current bounding box.center)},scale=.7]
    \draw[black] (0,0) circle (1);
    \draw (-1,0) .. controls (0,0) .. (0,-1);
    \draw (1,0) .. controls (0,0) .. (0,1);
    \node at (-1,0) [circle,fill,black,inner sep=1.pt]{};
    \node at (1,0) [circle,fill,black,inner sep=1.pt]{};
    \node at (0,1) [circle,fill,black,inner sep=1.pt]{};
    \node at (0,-1) [circle,fill,black,inner sep=1.pt]{};
\end{tikzpicture}
+
\begin{tikzpicture}[baseline={([yshift=-0.1cm]current bounding box.center)},scale=.7]
    \draw[black] (0,0) circle (1);
    \draw (-1,0) .. controls (0,0) .. (1,0);
    \draw (0,-1) .. controls (0,0) .. (0,1);
    \node at (-1,0) [circle,fill,black,inner sep=1.pt]{};
    \node at (1,0) [circle,fill,black,inner sep=1.pt]{};
    \node at (0,1) [circle,fill,black,inner sep=1.pt]{};
    \node at (0,-1) [circle,fill,black,inner sep=1.pt]{};
\end{tikzpicture}
= 2 + q^2~.
\ee
The above combinatorial procedure naturally leads to an inner-product structure on the Hilbert space spanned by the successive actions of $H$ on the empty state:
\be \label{eq:H-state-boundary}
\mathcal{H}_{0} = \mathrm{span}\{\, H^{k} |0\rangle \;|\; k = 0, 1, 2, \dots \}~.
\ee
One may visualize the states in~\eqref{eq:H-state-boundary} as living on the boundary of a sliced circle with a specified number of $H$-insertions.  
The action of $H$ on such a state simply corresponds to adding a new insertion on the boundary:
\be
H ~ 
\begin{tikzpicture}[baseline={([yshift=-0.1cm]current bounding box.center)},scale=0.35]
    \draw[thick] (0,-0.5) arc (0:-180:2); 
    \node at (-1,-2.23) [circle,fill,black,inner sep=1.2pt]{};
    \node at (-2,-2.5) [circle,fill,black,inner sep=1.2pt]{};
    \node at (-3,-2.23) [circle,fill,black,inner sep=1.2pt]{};
\end{tikzpicture}
=
\begin{tikzpicture}[baseline={([yshift=-0.1cm]current bounding box.center)},scale=0.35]
    \draw[thick] (0,-0.5) arc (0:-180:2); 
    \node at (-1,-2.23) [circle,fill,black,inner sep=1.2pt]{};
    \node at (-2,-2.5) [circle,fill,black,inner sep=1.2pt]{};
    \node at (-3,-2.23) [circle,fill,black,inner sep=1.2pt]{};
    \node at (-3.73,-1.5) [circle,fill,black,inner sep=1.2pt]{};
\end{tikzpicture}~.
\ee

According to the chord rules, the states in~\eqref{eq:H-state-boundary} are not orthogonal: their inner products involve summing over all $q$-weighted Wick contractions.  
It is therefore convenient to construct an orthonormal basis for $\mathcal{H}_0$, and two such bases will play a central role in the subsequent discussion.

One convenient basis for $\mathcal{H}_0$ is given by the chord number states, denoted by $|n\rangle$.  
These states are obtained by applying Schmidt orthogonalization to~\eqref{eq:H-state-boundary}\footnote{See also~\cite{xu2024von} for a normal-ordering prescription used in the construction of these states.}.  
They can be interpreted as states living on a spatial slice of an open chord diagram~\cite{Lin:2022rbf}, characterized by a fixed number of open chords ending on the slice.  

The action of the Hamiltonian chord operator $H$—defined as adding a new marked point on the boundary circle—has two effects on such a state:  
it either (i) creates a new open chord sourced by the new boundary insertion, or (ii) annihilates an existing open chord by connecting it to the new insertion, thereby forming a closed chord.  
To capture these two processes, it is useful to introduce chord creation and annihilation operators, $\mathfrak{a}^\dagger$ and $\mathfrak{a}$, whose diagrammatic actions are illustrated as:
\be \label{eq:add}
\mathfrak{a}^\dagger ~
\begin{tikzpicture}[baseline={([yshift=-0.1cm]current bounding box.center)},scale=0.35]
    \draw[thick] (0,-0.5) arc (0:-180: 2);
    \draw (-1,-0.5) -- (-1,-2.23);
    \draw (-2,-0.5) -- (-2,-2.5);
    \draw (-3,-0.5) -- (-3,-2.23);  
    \node at (-1,-2.23) [circle,fill,black,inner sep=1.2pt]{};
    \node at (-2,-2.5) [circle,fill,black,inner sep=1.2pt]{};
    \node at (-3,-2.23) [circle,fill,black,inner sep=1.2pt]{};
\end{tikzpicture}
=
\begin{tikzpicture}[baseline={([yshift=-0.1cm]current bounding box.center)},scale=0.35]
    \draw[thick] (0,-0.5) arc (0:-180: 2);
    \draw (-1,-0.5) -- (-1,-2.23);
    \draw (-2,-0.5) -- (-2,-2.5);
    \draw (-3,-0.5) -- (-3,-2.23);  
    \draw (-3.5,-0.5) -- (-3.5,-1.8);
    \node at (-1,-2.23) [circle,fill,black,inner sep=1.2pt]{};
    \node at (-2,-2.5) [circle,fill,black,inner sep=1.2pt]{};
    \node at (-3,-2.23) [circle,fill,black,inner sep=1.2pt]{};
    \node at (-3.5,-1.8) [circle,fill,black,inner sep=1.2pt]{};
\end{tikzpicture}~,
\qquad
\mathfrak{a}~
\begin{tikzpicture}[baseline={([yshift=-0.1cm]current bounding box.center)},scale=0.35]
    \draw[thick] (0,-0.5) arc (0:-180: 2);
    \draw (-1,-0.5) -- (-1,-2.23);
    \draw (-2,-0.5) -- (-2,-2.5);
    \draw (-3,-0.5) -- (-3,-2.23);  
    \node at (-1,-2.23) [circle,fill,black,inner sep=1.pt]{};
    \node at (-2,-2.5) [circle,fill,black,inner sep=1.pt]{};
    \node at (-3,-2.23) [circle,fill,black,inner sep=1.pt]{};
\end{tikzpicture}
=
\begin{tikzpicture}[baseline={([yshift=-0.1cm]current bounding box.center)},scale=0.35]
    \draw[thick] (0,-0.5) arc (0:-180: 2);
    \draw (-1,-0.5) -- (-1,-2.23);
    \draw (-2,-0.5) -- (-2,-2.5);
    \draw (-3.5,-1.8) .. controls (-3,-1.3) .. (-3,-2.23);
    \node at (-1,-2.23) [circle,fill,black,inner sep=1.pt]{};
    \node at (-2,-2.5) [circle,fill,black,inner sep=1.pt]{};
    \node at (-3,-2.23) [circle,fill,black,inner sep=1.pt]{};
    \node at (-3.5,-1.8) [circle,fill,black,inner sep=1.pt]{};
\end{tikzpicture}
+
\begin{tikzpicture}[baseline={([yshift=-0.1cm]current bounding box.center)},scale=0.35]
    \draw[thick] (0,-0.5) arc (0:-180: 2);
    \draw (-1,-0.5) -- (-1,-2.23);
    \draw (-3,-0.5) -- (-3,-2.23);  
    \draw (-2,-2.5) .. controls (-2.5,-1.3) and (-3,-1.3) .. (-3.5,-1.8);
    \node at (-1,-2.23) [circle,fill,black,inner sep=1.pt]{};  \node at (-3.5,-1.8) [circle,fill,black,inner sep=1.pt]{}; 
    \node at (-2,-2.5) [circle,fill,black,inner sep=1.pt]{};
    \node at (-3,-2.23) [circle,fill,black,inner sep=1.pt]{};
\end{tikzpicture}  +\begin{tikzpicture}[baseline={([yshift=-0.1cm]current bounding box.center)},scale=0.35]
    \draw[thick] (0,-0.5) arc (0:-180: 2);
    \draw (-2,-0.5) -- (-2,-2.5);
    \draw (-3,-0.5) -- (-3,-2.23);  
    \draw (-1,-2.23) .. controls (-2,-1.3) and (-3,-1.3)  .. (-3.5,-1.8);
    \node at (-1,-2.23) [circle,fill,black,inner sep=1.pt]{};  \node at (-3.5,-1.8) [circle,fill,black,inner sep=1.pt]{}; 
    \node at (-2,-2.5) [circle,fill,black,inner sep=1.pt]{};
    \node at (-3,-2.23) [circle,fill,black,inner sep=1.pt]{};
\end{tikzpicture},
\ee
where the annihilation operator sums over all possible ways of forming a closed chord by attaching one of the existing open chords to the newly added boundary point.  
Each configuration carries an appropriate $q$-weight determined by the number of crossings introduced in this process.  

The illustration corresponds to the case $n=3$, showing the action of the ladder operators on a state with a fixed chord number:
\be
\mathfrak{a}^\dagger |n\rangle = |n+1\rangle~,
\qquad
\mathfrak{a} |n\rangle = [n]_{q^{2}}\, |n-1\rangle~,
\ee
where $[n]_{q^2} = 1 + q^2 + \dots + q^{2(n-1)}$ is the $q$-integer.  
These operators satisfy the $q$-deformed canonical commutation relation
\be \label{eq:number-def}
[\mathfrak{a}, \mathfrak{a}^\dagger]_{q^2}
\equiv \mathfrak{a}\mathfrak{a}^\dagger - q^2 \mathfrak{a}^\dagger \mathfrak{a}
= 1~,\quad [\mathfrak{a},\mathfrak{a}^{\da}]=q^{2\hat{n}},\, \text{ where }\,\hat{n}|n)=n|n).
\ee
from which one finds
$\langle n|m\rangle = \delta_{mn}\,[m]_{q^2}!$,  
where $[n]_q! = \prod_{j=1}^{n} [j]_q$ denotes the $q$-factorial.  
The inner product on $\mathcal{H}_0$ ensures that $\mathfrak{a}$ and $\mathfrak{a}^\dagger$ are Hermitian conjugates.  

It is convenient to introduce normalized chord-number states $|n) = ([n]_{q^2}!)^{-1/2} |n\rangle$ and rescaled ladder operators $a, a^\dagger$ with
\be
a^\dagger |n) = \sqrt{1 - q^{2(n+1)}}\,|n+1)~, 
\qquad 
a |n) = \sqrt{1 - q^{2n}}\,|n-1)~, \quad (m|n)=\delta_{mn}~.
\ee
In this normalization, the chord Hamiltonian takes the simple form
\be \label{eq:zero-H-def}
H = \mathfrak{a}^\dagger + \mathfrak{a}
= \frac{a^\dagger + a}{\sqrt{1 -q^2}}~.
\ee
It is clear that $H$ is a Hermitian operator, and therefore one can perform a spectral decomposition that resolves $\mathcal{H}_0$ into its energy eigenstates.  
An eigenstate of $H$ with energy label $\theta$ was constructed in~\cite{Berkooz:2018jqr} as
\be
|\theta) = \sum_{n=0}^{\infty} 
\frac{H_n(\cos\theta\,|\,q^2)}{\sqrt{(q^2;q^2)_n}}\, |n)~,
\ee
where $H_n(\cos\theta\,|\,q^2)$ denotes the $n$-th $q$-Hermite polynomial and $(x;q)_n$ is the $q$-Pochhammer symbol.  
The coefficients are chosen such that the energy eigenstates form an orthonormal basis:
\be
(\theta_1|\theta_2)
= \sum_{n=0}^{\infty}
\frac{H_n(\cos\theta_1\,|\,q^2)\,H_n(\cos\theta_2\,|\,q^2)}{(q^2;q^2)_n}
= \frac{\delta(\theta_1-\theta_2)}{\mu(\theta)}~,
\qquad 
\mu(\theta)= \frac{(e^{\pm 2i\theta},q^2;q^2)_\infty}{2\pi}~,
\ee
where $\mu(\theta)$ defines the energy density, compactly supported on $\theta\in[0,\pi]$. We have used the convention $(x_1,\ldots, x_k;q)_n=\prod_{i=1}^k(x_i;q)_n$ in defining $\mu(\theta)$. The spectral decomposition of $\mathcal{H}_0$ then takes the form
\be
\mathbf{1}_{\mathcal{H}_0}
= \int_{0}^{\pi} \! \dd\mu(\theta)\, |\theta)(\theta|
= \sum_{n=0}^{\infty} |n)(n|~,
\ee
where we use the short-hand notation $\dd\mu(\theta) = \mu(\theta)\,\dd\theta$.  

The construction of $\mathcal{H}_0$ and its spectral resolution provides an analytic framework for solving the original moment problem~\eqref{eq:doublescaling}, reorganizing the $q$-weighted combinatorial structure into one with clear geometric and algebraic interpretation.  
This perspective becomes even more illuminating when matter is included, as it reveals the emergence of an underlying quantum group structure directly from the chord description.

\subsection{The Bulk Hilbert Space with Matter and Isometric Factorization}
We begin by reviewing the construction of the one-particle chord Hilbert space $\mathcal{H}_1$, corresponding to a single matter chord insertion. This space is spanned by states characterized by a fixed matter insertion of weight $\Delta > 0$, with $n_L$ and $n_R$ denoting the number of $H$-chords to the left and right of the matter chord, respectively:
\be
\mathcal{H}_{1} = \mathrm{span} \left\{ |\Delta, n_L, n_R\rangle \,\big|\, n_L, n_R \in \mathbb{Z}_{\geq 0} \right\}~.
\ee
Chord creation and annihilation operators act by adding or removing an $H$-chord to the left or right of the matter insertion. Explicitly, the raising operators act as:
\be
\mathfrak{a}_L^{\dagger} |\Delta; n_L, n_R\rangle = |\Delta; n_L + 1, n_R\rangle~, \quad
\mathfrak{a}_R^{\dagger} |\Delta; n_L, n_R\rangle = |\Delta; n_L, n_R + 1\rangle~,
\ee
while the lowering operators act according to:
\be \label{eq:annihilators-def}
\begin{aligned}
\mathfrak{a}_L |\Delta; n_L, n_R\rangle &= [n_L]_{q^2}\, |\Delta; n_L - 1, n_R\rangle + q^{2\Delta + 2 n_L} [n_R]_{q^2}\, |\Delta; n_L, n_R - 1\rangle~, \\
\mathfrak{a}_R |\Delta; n_L, n_R\rangle &= [n_R]_{q^2}\, |\Delta; n_L, n_R - 1\rangle + q^{2\Delta + 2n_R} [n_L]_{q^2}\, |\Delta; n_L - 1, n_R\rangle~.
\end{aligned}
\ee
The presence of matter modifies the rules for annihilating an existing $H$-chord in a chord diagram containing an $M$-chord,  
by explicitly accounting for its crossing pattern with the $M$-chord.   Specifically, annihilating an $H$-chord to the left of an $M$-chord does not produce an $M$–$H$ crossing,   whereas annihilating an $H$-chord to the right of the $M$-chord yields a closed chord that crosses all $M$-chords  
and all $H$-chords to its left. This is shown explicitly in the coefficients of the two terms in the first equation of \eqref{eq:annihilators-def}. 


The $q$-ladder operators satisfy the following $q$-deformed commutation relations:
\be \label{eq:commutator-1}
\begin{aligned}
[\mathfrak{a}_L, \mathfrak{a}_L^{\dagger}]_{q^2} &= [\mathfrak{a}_R, \mathfrak{a}_R^{\dagger}]_{q^2} = 1, \\
[\mathfrak{a}_L, \mathfrak{a}_R^{\dagger}] &= [\mathfrak{a}_R, \mathfrak{a}_L^{\dagger}] = q^{2\hat{n}_{\text{tot}}},
\end{aligned}
\ee
where $\hat{n}_{\text{tot}}$ is the total chord number operator, defined via $\hat{n}_{tot}|\Delta;m_L,m_R\rangle = (m_L+m_R+\Delta)|\Delta;m_L,m_R\rangle$, and the $q$-commutator is defined as $[A,B]_q\equiv AB-q BA$.  The inner product on $\mathcal{H}_1$ is defined so that $\mathfrak{a}_{L/R}$ and $\mathfrak{a}^{\dagger}_{L/R}$ are Hermitian conjugates. This leads to a recursive definition of the inner product:
\be \label{eq:recur-1}
\begin{split}
    J_{n_L,n_R;m_L,m_R}& =\langle \Delta; n_L, n_R | \Delta; m_L, m_R \rangle = \langle \Delta; n_L - 1, n_R | \mathfrak{a}_L | \Delta; m_L, m_R \rangle, \\ 
    &= [m_L]_{q^2} J_{n_L-1,n_R;m_L-1,m_R}+ q^{2\Delta + 2m_L} [m_R]_{q^2} J_{n_L-1,n_R;m_L,m_R-1}~.
\end{split}
\ee 
subject to the boundary conditions:
\be
\langle \Delta, m, 0 | \Delta, n, 0 \rangle = \langle \Delta, 0, m | \Delta, 0, n \rangle = \delta_{mn}[n]_{q^2}!~,
\ee
which reduces to the zero-particle inner product of $\mathcal{H}_0$, where all $H$-chords lie on the same side of the $M$-chord and do not intersect it. 

A solution to the recursion relation \eqref{eq:recur-1} was provided in \cite{Lin_2023} by mapping it to a random walk problem with $q$-deformed weights. The resulting expression is given as a sum over weighted lattice paths. For our purpose, the more relevant approach to solving \eqref{eq:recur-1} was introduced in \cite{xu2024von}, which constructs eigenstates of the left and right Hamiltonians defined by
\be
H_{L/R} = \mathfrak{a}_{L/R} + \mathfrak{a}^\dagger_{L/R}~,
\ee
subject to the commutation relations \eqref{eq:commutator-1}. One can then verify that these Hamiltonians commute, $[H_L, H_R] = 0$, therefore can be simultaneously diagonalized. Denoting the eigenstate as $|\Delta; \te_L,\te_R\ket$, it satisfies
\be
H_{L/R}|\Delta;\te_{L},\te_{R}\ket=E(\te_{L/R})|\Delta;\te_{L},\te_{R}\ket~,\quad E(\te_{L/R})=\frac{2\cos\te_{L/R}}{\sqrt{1-q^2}}~.
\ee
An explicit construction of the state $|\Delta; \theta_L, \theta_R\rangle$ is provided in \cite{Xu:2025chord}, along with a useful characterization of the one-particle Hilbert space $\mathcal{H}_1$ in terms of the doubled zero-particle Hilbert space $\mathcal{H}_0 \otimes \mathcal{H}_0$ (see also \cite{Okuyama:2024yya}). In particular, there exists a unitary intertwiner map $\Upsilon_\Delta: \mh_1\to \mh_0\otimes \mh_0$, which isometrically factorizes $\mh_1$. That is,
\be
\Upsilon_{\Delta}^\dagger \Upsilon_\Delta = \mathbf{1}_{\mh_1}~, \quad \Upsilon_\Delta \Upsilon^{\dagger}_\Delta = \mathbf{1}_{\mh_0\otimes \mh_0}~. 
\ee
The intertwiner $\Upsilon_\Delta$ is characterized as converting the one-particle Hamiltonians $H_{L/R}$ to the zero-particle Hamiltonians $h_{L/R}$ acting on the left/right copy respectively:
\be
\Upsilon_\Delta \circ H_{L/R} = h_{L/R} \circ \Upsilon_\Delta~.
\ee
Here we used $\circ$ to denote composition of maps, since $\Upsilon_\Delta$ maps states to a different Hilbert space. To distinguish it from the left/right one-particle Hamiltonian, we adopt the notation that \( h_{L/R} \) denotes the zero-particle Hamiltonian defined in~\eqref{eq:zero-H-def}, acting on the left/right copy of \( \mathcal{H}_0 \), respectively.  The above property ensures that $\Upsilon_\Delta$ maps eigenstates of $H_{L/R}$ to eigenstates of $h_{L/R}$.  Specifically, we have:
\be
\Upsilon_{\Del}|\Delta;\te_{L},\te_{R}\ket=\gamma_{\Del}(\te_{L},\te_{R})|\te_{L},\te_{R})~,\quad\Upsilon_{\Del}^{\da}|\te_{L},\te_{R})=\frac{1}{\gamma_{\Del}(\te_{L},\te_{R})}|\Delta;\te_{L},\te_{R}\ket~,
\ee
where states written with round brackets live in (products of) $\mathcal{H}_0$, and we use the shorthand notation $|\theta_L,\theta_R)\equiv|\theta_L)\otimes|\theta_R)$.  The matter density $\gamma_{\Delta}(\te_L,\te_R)$ is defined as
\be
\gamma_{\Del}(\theta_L,\theta_R)\equiv N_{q^2}^{1/2}\sqrt{\frac{\Gamma_{q^2}(\Del\pm\frac{i\te_{L}}{\lambda}\pm\frac{i\te_{R}}{\lambda})}{\Gamma_{q^2}(2\Del)}}~, \quad \text{ where } \quad \lambda \equiv -\log(q)~,
\ee
with $N_q$ being only a function of $q$, and $\pm$ in the $q$-gamma function\footnote{We refer the reader to Appendix~\ref{app:review} for a review of the definition and key properties of the $q$-gamma function.} means that we take the product of the functions with both $+$ and $-$ inserted. The normalization is chosen so that the inner product of two one-particle fixed energy states agrees with the two-point correlation function with fixed boundary energy:
\be \label{eq:matter-density-def}
\begin{aligned}
\left\langle\Delta ; \theta_L, \theta_R \mid \Delta ; \phi_L, \phi_R\right\rangle & =\left\langle\Delta ; \theta_L, \theta_R\right| \Upsilon_{\Delta}^{\dagger} \Upsilon_{\Delta}\left|\Delta ; \phi_L, \phi_R\right\rangle  =\gamma_{\Delta}^2\left(\theta_L, \theta_R\right)\left(\theta_L, \theta_R \mid \phi_L, \phi_R\right) \\
& =N_{q^2} \frac{\Gamma_{q^2}\left(\Delta \pm \frac{i \theta_L}{\lambda} \pm \frac{i \theta_R}{\lambda}\right)}{\Gamma_{q^2}(2 \Delta)} \frac{\delta\left(\theta_L-\phi_L\right) \delta\left(\theta_R-\phi_R\right)}{\mu\left(\theta_L\right) \mu\left(\theta_R\right)}~.
\end{aligned}
\ee 
In the first equality, we insert an identity in $\mathcal{H}_1$, and in the second equality, we use the definition of $\Upsilon_\Delta$ to express the inner product in terms of the inner product of states in $\mathcal{H}_0 \otimes \mathcal{H}_0$. 

The image of fixed chord number states under $\Upsilon_\Delta$ is more involved, and is solved in \cite{xu2024von}. It can be expressed as:
\be \label{eq:intertwiner-chord}
\Upsilon_{\Del}|\Delta;m_{L},m_{R}\ket=\int_{0}^{\pi}\dd\mu(\te_{L})\dd\mu(\te_{R}) \gamma_\Delta(\te_L,\te_R) |\te_{L},\te_{R})(\te_{L},\te_{R}|(q^{2\Delta} a_L a_R;q^2)_\infty|m_{L},m_{R})~,
\ee
where the two-sided operator $(q^{2\Delta} a_L a_R;q^2)_\infty$ is understood in terms of the expansion:
\be
(q^{2\Del}a_{L}a_{R};q^2)_{\infty}=\sum_{k=0}^{\infty}\frac{(-1)^{k}q^{2\binom{k}{2}+2 k\Del}}{(q^2;q^2)_{k}}a_{L}^{k}a_{R}^{k}~.
\ee
Here, $a_{L/R}$ denote the $q$-deformed annihilation operators acting on the left and right copies of $\mathcal{H}_0$, respectively. The series above truncates when acting on any fixed chord number state, and thus defines a bounded operator on $\mathcal{H}_0 \otimes \mathcal{H}_0$. Utilizing the relation \eqref{eq:intertwiner-chord}, the inner product \eqref{eq:recur-1} can be brought into the following explicit form:
\be \label{eq:J-sol}
\begin{split}
    \bra & \Delta; m_L , m_R | \Delta ; n_L, n_R \ket  = \bra \Delta; m_L , m_R | \Upsilon^{ \dagger}_\Delta\Upsilon_{\Delta}|\Delta ; n_L, n_R \ket \\ 
    &= \int \dd \mu(\te_L) \dd \mu(\te_R) N_{q^2} \frac{\Gamma_{q^2}\left(\Delta \pm \frac{i \theta_L}{\lambda} \pm \frac{i \theta_R}{\lambda}\right)}{\Gamma_{q^2}(2 \Delta)}  \phi^{\Delta*}_{E_L,E_R}(m_L,m_R) \phi^\Delta_{E_L,E_R} (n_L,n_R)~,
\end{split}
\ee
where the one-particle wavefunctions with fixed boundary energy $E_{L/R}=E(\te_{L/R})$ is introduced~\cite{xu2024von,Okuyama:2024gsn}:
\be
\phi_{E_{L}E_{R}}^{\Del}(m_{L},m_{R})\equiv (\te_{L},\te_{R}|(q^{2\Delta}a_L a_R;q^2)_\infty|m_{L},m_{R})~.
\ee 
The expression \eqref{eq:J-sol} solves \eqref{eq:recur-1} in a manner reminiscent of JT gravity coupled to $\mathfrak{sl}_2$ matter on the Euclidean disk, where the matter density is interpreted as the matrix element $|\mathcal{O}_{E_L E_R}|^2$ of a primary operator with conformal weight $\Delta$, see~\cite{jafferis2023,Kolchmeyer:2023gwa,Iliesiu:2024cnh} for related discussions.

\subsection{A Curious Recursion}

Realizing the equivalence between $\mathcal{H}_1$ and $\mathcal{H}_0 \otimes \mathcal{H}_0$, we now study states in the latter Hilbert space, which can naturally be interpreted as the space of bounded operators acting on $\mathcal{H}_0$\footnote{This equivalence was explored in~\cite{Okuyama:2024yya}.  In our discussion, we reformulate it in terms of an intertwiner to clearly distinguish states belonging to the two different Hilbert spaces.
}. A state of particular interest is the following:
\begin{equation} \label{eq:state0-def}
|q^{2\Delta \hat{n}}) \equiv \sum_{n=0}^{\infty} q^{2\Delta n} |n,n)~,
\end{equation}
where the number operator $\hat{n}$ is defined in \eqref{eq:number-def}. The overlap with energy eigenstates produces the matter density\footnote{We construct the states by acting with the creation operators from the left; an analogous construction holds for operators acting from the right. As will become clear in the subsequent discussion, it is the two-sided combination of these ladder operators that gives rise to the quantum group structure.}:
\begin{equation}
(\theta_L, \theta_R | q^{2\Delta \hat{n}}) = \sum_{n=0}^{\infty} \frac{q^{2\Delta n} H_n(\cos\theta_L | q^2) H_n(\cos\theta_R | q^2)}{(q^2;q^2)_n} = \frac{(q^{4\Delta};q^2)_{\infty}}{(q^{2\Delta}e^{i(\pm \theta_L\pm \theta_R)};q^2)}=\gamma_{\Delta}(\te_L,\te_R)^2~.
\end{equation}
The second equality follows from the bivariate generating function of the $q$-Hermite polynomials and coincides with the matter density defined in~\eqref{eq:matter-density-def}, after expressing the $q$-Pochhammer symbols in terms of the corresponding $q$-gamma functions using~\eqref{eq:qGamma-def}. Other states can be generated by acting with the ladder operators on the base state. In particular, we define the family of states:
\begin{equation} \label{eq:statem-def}
\widetilde{|\Delta;m]} \equiv 
\sqrt{\frac{(q^{4\Delta};q^2)_m}{(q^2;q^2)_m}}\,
a^{\dagger m}_L |q^{2\Delta \hat{n}})~,\quad m=0,1,2,\dots,
\end{equation}
Here, the tilde is used to distinguish these states from their normalized counterparts, which will be constructed below. Although these states are not normalized, they are convenient to work with due to their simple definition in \eqref{eq:statem-def}. Notably, the states defined in~\eqref{eq:statem-def} form an entangled basis in the doubled Hilbert space, allowing the action of commuting sets of left and right ladder operators. This formulation makes the underlying quantum group structure, arising from two-sided combinations of these operators, as well as the action of the corresponding generators on these states, particularly transparent.

In addition, states defined in \eqref{eq:statem-def} satisfy the following recursion relation:
\begin{equation} \label{eq:eigenE}
\begin{aligned}
(\theta_L, \theta_R |E(\theta_L, \theta_R) \widetilde{|\Delta, m ]} & =a_{m+1} (\theta_L, \theta_R \widetilde{| \Delta, m+1]}+a_{m-1} (\theta_L, \theta_R \widetilde{|\Delta, m-1] } \\
& +b_m (\theta_L, \theta_R \widetilde{| \Delta, m]}~,
\end{aligned}
\end{equation}
which we derive in Appendix~\ref{app:recursion}. The coefficients and the `two-sided' energy $E(\theta_L,\theta_R)$ are defined as:
\begin{equation} \label{eq:two-sidedenergy}
\begin{aligned}
a_m & = \frac{\sqrt{(1 - q^{2m})(1 - q^{4\Delta+2m - 2})}}{q^{-1} - q}~, \\
b_m & = \frac{2\cos\theta_R}{q^{-1} - q}(q^{2\Delta + 2m} - 1)~, \\
E(\theta_L, \theta_R) & = \frac{2\cos\theta_L - 2\cos\theta_R}{q^{-1} - q}~.
\end{aligned}
\end{equation}
This recursion defines the so-called Al-Salam Chihara polynomials\footnote{See~\cite{szabłowski2013} for a review of the definitions and properties of these polynomials, and~\cite{Berkooz:2018jqr} for their appearance in the context of the double-scaled SYK model.} with appropriate normalization, and we show in the next section how these states furnish the positive discrete series representation of $\mathcal{U}_q(\mathfrak{su}(1,1))$. Explicitly, the overlap with fixed energy states is given by:
\begin{equation} \label{eq:components}
(\theta_L, \theta_R \widetilde{| \Delta; m]} = (\theta_L, \theta_R | q^{2\Delta \hat{n}}) \frac{Q_m(\cos\theta_L | q^{2\Delta} e^{\pm i\theta_R}; q^2)}{\sqrt{(q^{4\Delta}, q^2; q^2)_m}}~,
\end{equation}
and we refer the reader to Appendix C of \cite{Xu:2024gfm} for a detailed derivation. 

Despite their simple definition in terms of number and ladder operators, the states defined in \eqref{eq:statem-def} are not orthonormal with respect to the chord inner product\footnote{Here, we refer to the inner product in $\mh_0\otimes \mh_0$, which is equivalent to the one in $\mh_1$ based on the isometry established in earlier discussion.}, and are not directly related to states in $\mathcal{H}_1$. The one-particle state $|\Delta;0,0\rangle$ and $|q^{2\Delta \hat{n}})$ are related by: 
\begin{equation} \label{eq:def-Vd}
\Upsilon_\Delta |\Delta;0,0\rangle = \hat{V}_\Delta |q^{2\Delta \hat{n}})~, \quad \hat{V}_\Delta = \int_0^{\pi} d\mu(\theta_L) d\mu(\theta_R) \frac{1}{\gamma_\Delta(\theta_L, \theta_R)} |\theta_L, \theta_R)(\theta_L, \theta_R|~.
\end{equation}
It is easy to see that $\Vd=\Vd^\dagger$. If we introduce:
\begin{equation} \label{eq:statem-def2}
|\Delta;m] \equiv \hat{V}_\Delta \widetilde{|\Delta;m]}~,
\end{equation}
one can show that the states $\{|\Delta;m]\}$ are orthonormal with respect to the inner product in $\mathcal{H}_0 \otimes \mathcal{H}_0$, while also satisfying the recursion relation \eqref{eq:eigenE}. We show the orthonormality condition in Appendix~\ref{app:orthonormal}.

A curious feature of the states \eqref{eq:statem-def2} is that they are orthogonal under a reduced inner product on $\mathcal{H}_0 \otimes \mathcal{H}_0$ that depends only on $\theta_L$, and is independent of $\theta_R$. Specifically, one can verify:
\begin{equation}
\int d\mu(\theta_L) [\Delta;m| \theta_L, \theta_R)(\theta_L, \theta_R |\Delta;n] = \delta_{mn}~,
\end{equation}
which follows from substituting \eqref{eq:components} and \eqref{eq:def-Vd}, and using the standard orthogonality relation of the Al-Salam–Chihara polynomials:
\begin{equation} \label{eq:orthogonality-Q}
\int d\mu(\theta_L) (\theta_L, \theta_R | q^{2\Delta \hat{n}}) \frac{Q_m(\cos\theta_L | q^{2\Delta} e^{\pm i\theta_R}, q^2) Q_n(\cos\theta_L | q^{2\Delta} e^{\pm i\theta_R}, q^2)}{(q^{4\Delta}, q^2; q^2)_m} = \delta_{mn}~,
\end{equation}
which holds for any $\theta_R \in [0, \pi]$. This property enables a decomposition of $\mathcal{H}_0 \otimes \mathcal{H}_0$, and consequently $\mathcal{H}_1$, into a direct sum of subspaces spanned by $|\Delta;m]$. We elaborate further on this decomposition after establishing how these states furnish irreducible representations of $\mathcal{U}_q(\mathfrak{su}(1,1))$ in the next section.

\section{The Quantum Group Structure and its Representation} \label{sec:qg-structure}

In this section, we show how the quantum group $\mathcal{U}_q(\mathfrak{su}(1,1))$ acts on the one-particle sector $\mathcal{H}_1$.   Specifically, we show that the recursion relation~\eqref{eq:eigenE} realizes the positive discrete series representation of $\mathcal{U}_q(\mathfrak{su}(1,1))$, and we construct its generators explicitly in terms of the chord algebra generated by the left and right chord ladder operators.  
We then perform a spectral decomposition of $\mathcal{H}_1$ with respect to a particular self-adjoint operator built from these generators, allowing for a natural identification between fixed-energy states in $\mathcal{H}_1$ and the corresponding eigenstates of this operator.  
This leads to a decomposition of $\mathcal{H}_1$ into a tensor product of positive discrete series representations.   Building on this structure, we extend the analysis to multi-particle states, showing that they can be systematically obtained through the fusion of one-particle representations.  
Finally, we apply the resulting fusion rules to derive the chord-level Yang--Baxter relation.

\subsection{Discrete Series Representation of $\mathcal{U}_q(\mathfrak{su}(1,1))$}
We begin this section by reviewing the quantum group $\uqsu$ and its positive discrete series representations. We start with the quantum group $\uqsl$, which is the universal enveloping algebra generated by four elements $K$, $K^{-1}$, $E$, and $F$, which satisfy the following commutation relations:
\be \label{eq:defining}
KK^{-1}=1=K^{-1}K~,\quad KE=qEK~,\quad KF=q^{-1}FK~,\quad EF-FE=\frac{K^{2}-K^{-2}}{q-q^{-1}}~.
\ee
\justifying
Note that this differs from the Drinfeld--Jimbo representation adopted in~\cite{Mertens:2025qg1}, which is an equivalent formulation better suited for discussing $\uqslr$, whereas the representation~\eqref{eq:defining} is more natural for $\uqsu$. The distinction between the two will be clarified in the subsequent discussion, and in Appendix~\ref{app:transform} we explicitly show how to transform the representation~\eqref{eq:defining} into the Drinfeld--Jimbo form. By parameterizing $K$ as $K=q^{2H}$, the algebra takes a more familiar form:
\be
[H,E]=E~,\quad[H,F]=-F~,\quad[E,F]=\frac{q^{2H}-q^{-2H}}{q-q^{-1}}~,
\ee
which clearly reduces to the classical $\mathfrak{sl}_2$ Lie algebra commutation relations in the $q\to 1$ limit. The Casimir operator $\Omega$ of the algebra is defined by
\be 
\Omega=\frac{q^{-1}K^{2}+qK^{-2}-2}{\left(q^{-1}-q\right)^{2}}+EF=\frac{q^{-1}K^{-2}+qK^{2}-2}{\left(q^{-1}-q\right)^{2}}+FE~.
\ee
The associative algebra generated by the four generators in \eqref{eq:defining}, together with the coproduct structure, which we introduce later, forms a so-called Hopf algebra. There are several inequivalent $*$-structures that can be associated with it. The one relevant for defining $\uqsu$ is\footnote{For alternative $*$-structures that give rise to $\uqslr$ and $\mathcal{U}_q(\mathfrak{su}(2))$, see~\cite{Mertens:2025qg1}. 
Although these $*$-structures are inequivalent for general $q$, they all coincide in the $q \to 1$ limit, reducing to the standard $*$-structure of the universal enveloping algebra $U(\mathfrak{sl}_2)$ generated by the three classical $\mathfrak{sl}_2$ generators.}:
\be \label{eq:star-structure}
*K=K~,\quad*K^{-1}=K^{-1}~,\quad*E=-F~,\quad*F=-E~.
\ee

We now consider the positive discrete series representation $\mathcal{D}^{+}_\Delta$ of $\uqsu$ with weight $\Delta$. The basis states of the representation are denoted by $|\Delta;n]$.\footnote{These will later be identified with states introduced in \eqref{eq:statem-def}, although they currently refer to states in $\mathcal{D}^{+}_\Delta$.} The action of the generators on $\mathcal{D}^{+}_\Delta$ is given by \cite{koelink1996}:
\begin{align}\label{eq:discrete-rep}
K|\Delta;n] &= q^{\Delta+n}|\Delta;n]~,\quad K^{-1}|\Delta;n]=q^{-\Delta-n}|\Delta;n]~,\\
\left(q^{-1}-q\right)E|\Delta;n] &= q^{-\frac{1}{2}-\Delta-n}\sqrt{\left(1-q^{2n+2}\right)\left(1-q^{4\Delta+2n}\right)}|\Delta,n+1]~,\\
\left(q^{-1}-q\right)F|\Delta;n] &= -q^{\frac{1}{2}-\Delta-n}\sqrt{\left(1-q^{2n}\right)\left(1-q^{4\Delta+2n-2}\right)}|\Delta,n-1]~,\\
\left(q^{-1}-q\right)^{2}\Omega|\Delta;n] &= \left(q^{2\Delta-1}+q^{-2\Delta+1}-2\right)|\Delta;n]~,
\end{align}
with 
\be
\mathcal{D}^{+}_\Delta \equiv \text{span}\{\,|\Delta;m] \,|\, m=0,1,2\dots \}~,\quad [\Delta;m|\Delta;n]=\delta_{mn}~.
\ee
Note that the above representation reduces to the standard positive discrete series representation of $\mathfrak{sl}_2$ in the limit $q \to 1$.

The $*$-structure introduces an intrinsic notion of self-adjointness for operators in $\uqsu$, which allows a spectral decomposition of a given irreducible representation. To see how this works with $\mathcal{D}^{+}_\Delta$,  we introduce a one-parameter family of operators $Y_\phi K$, parameterized by $\phi\in [0,\pi]$, through the assignment:
\be \label{eq:YK-def}
Y_{\phi} K \equiv q^{1/2}E K -q^{-1/2}F K+\frac{2\cos\phi}{q^{-1}-q}\left(K^2 -1\right)~.
\ee
It is easy to check that the operator $Y_\phi K$ satisfies $*(Y_\phi K) = Y_\phi K$ for any $\phi\in[0,\pi]$.  The eigenstates of the operator $Y_\phi K$ can be constructed from the basis $\{|\Delta; n] \mid n=0,1,2,\dots\}$ as
\begin{equation} \label{eq:eigenstate-2}
|\Delta; \theta, \phi\rangle = \left( (\theta, \phi|q^{2\Delta \hat{n}}) \sum_{n=0}^{\infty} \frac{Q_n(\cos\theta \mid q^{2\Delta}e^{\pm i\phi}, q^2)}{\sqrt{(q^2, q^{4\Delta}; q^2)_n}} |\Delta; n] \right) \otimes |\phi)~,
\end{equation}
where we have introduced the factor $|\phi)$ to keep track of the $\phi$-dependence in $|\Delta; \theta, \phi\rangle$ as an eigenstate of $Y_\phi K$. The normalization of the state is chosen so that one can express states with given $n$ on the right-hand side as:
\be \label{eq:reverse}
|\Delta,n]\otimes|\phi)=\int_{0}^{\pi}\dd\mu(\te)\frac{Q_{n}\left(\cos\te \mid q^{2\Del}e^{\pm i\phi};q^{2}\right)}{\sqrt{\left(q^{4\Del},q^{2};q^{2}\right)_{n}}}|\Delta;\te,\phi\ket~,
\ee
as follows from the orthogonality relation \eqref{eq:orthogonality-Q}. Note that we have intentionally used the same notation for the state in the left-hand side of \eqref{eq:eigenstate-2} as for the fixed-energy eigenstate in $\mathcal{H}_1$, since one can verify that the state defined in \eqref{eq:eigenstate-2} has an inner product that agrees with the one in $\mathcal{H}_1$:
\begin{equation} \label{eq:inner-product3}
\begin{aligned}
\langle \Delta; \theta, \phi | \Delta; \theta', \phi' \rangle &= \frac{\delta(\phi - \phi')}{\mu(\phi)} (\theta, \phi | q^{2\Delta \hat{n}})(\theta', \phi' | q^{2\Delta \hat{n}}) \\
&\quad \times \sum_{n=0}^{\infty} \frac{Q_n(\cos\theta \mid q^{2\Delta}e^{\pm i\phi}; q^2) Q_n(\cos\theta' \mid q^{2\Delta}e^{\pm i\phi}; q^2)}{(q^{4\Delta}, q^2; q^2)_n} \\
&= (\theta, \phi | q^{2\Delta \hat{n}}) \frac{\delta(\phi - \phi') \delta(\theta - \theta')}{\mu(\phi) \mu(\theta)}~,
\end{aligned}
\end{equation}
where in the second equality we used the completeness relation of the Al-Salam–Chihara polynomials. This validates the identification of the right-hand side of \eqref{eq:eigenstate-2} with the one-particle state $|\Delta; \theta, \phi\rangle$ in $\mathcal{H}_1$.

Denoting the coefficients in the summand of \eqref{eq:eigenstate-2} by $f_n$, we find the eigenvalue equation of $Y_\phi K$
\be \label{eq:E2-def}
Y_{\phi}K|\Delta;\theta,\phi\rangle=E(\theta,\phi)|\Delta;\theta,\phi\rangle~,\quad E(\theta,\phi)=\frac{2\cos\theta-2\cos\phi}{q^{-1}-q}~,
\ee
translates to a recursion relation for the coefficients $f_n$ as:
\be
\begin{aligned}
Y_{\phi}K|\Delta;\theta,\phi\rangle & =\left( \sum_{n=0}^{\infty}f_{n}\left(a_{n+1}|\Delta;n+1]+a_{n}|\Delta;n-1]+b_{n}|\Delta;n]\right)\right) \otimes |\phi)\\
&=\sum_{n=0}^{\infty}\left(a_{n}f_{n-1}+a_{n+1}f_{n+1}+b_{n}f_{n}\right)|\Delta;n]\otimes |\phi)~,
\end{aligned}
\ee
where the coefficients $a_n$ and $b_n$ are as defined in \eqref{eq:eigenE}. These results imply that, for each fixed value of \(\phi\), we obtain a copy of the positive discrete series representation \(\mathcal{D}^{+}_{\Delta}\) of the quantum group $\uqsu$. Consequently, the one-particle chord Hilbert space \(\mathcal{H}_{1}\) can be viewed as a direct integral of such representations labeled by \(\phi\), with measure \(\mathrm{d}\mu(\phi)\). In this way, integrating over \(\phi\) in the spectral decomposition associated with \(Y_{\phi}K\) naturally leads to the following decomposition:
\be \label{eq:crossed-product}
\mh^{\Delta}_1 \simeq \mathcal{D}^{+}_\Delta \otimes  \ml^2 ([0,\pi], \dd \mu (\phi))~,
\ee
The explicit $\phi$-dependence in the second factor implies that, for different values of $\phi$, these are all equivalent $\mathcal{D}^{+}_\Delta$ representations, each weighted by the zero-particle energy density $\dd\mu(\phi)$.
The emergence of this decomposition is not coincidental; rather, it reflects the underlying quantum group structure present in the chord algebra generated by \( a_{L/R}, a^{\dagger}_{L/R}\)\footnote{Note that from \eqref{eq:number-def}, the left/right number operators $q^{2\hat{n}_{L/R}}$ are also contained in the algebra.}, which are $q$-ladder operators and their corresponding number operators acting on the left or right copy of the zero-particle Hilbert space in $\mh_0\otimes\mh_0$.  They satisfy the following commutation relation:
\be \label{eq:chord-commutation}
\begin{aligned}
{\left[a_L, a_L^{\dagger}\right]_{q^2} } & =\left[a_R, a_R^{\dagger}\right]_{q^2}=1-q^2~, \\
{\left[a_{L / R}, a_{R / L}\right] } & =\left[a_{L / R}, a_{R / L}^{\dagger}\right]=0~, \\
{\left[\hat{n}_L, a_L^{\dagger}\right] } & =a_L^{\dagger}, \quad\left[ \hat{n}_L, a_L\right]=-a_L~, \\
{\left[\hat{n}_R, a_R^{\dagger}\right] } & =a_R^{\dagger}, \quad\left[\hat{n}_R, a_R\right]=-a_R~, \\
{\left[\hat{n}_{L / R}, a_{R / L}\right] } & =\left[\hat{n}_{L / R}, a_{R / L}^{\dagger}\right]=0~.
\end{aligned}
\ee
This algebra contains the generators of the quantum group \( \uqsu \), under the following identification:
\be \label{eq:id}
\begin{aligned} 
K & = q^{\hat{N}}, \quad K^{-1} = q^{-\hat{N}}, \\
(q^{-1} - q) E & = q^{\frac{1}{2}} \left( q^{-\hat{N}} a_L^\dagger - q^{\hat{N}} a_R \right), \\
(q^{-1} - q) F & = - q^{-\frac{1}{2}} \left( q^{-\hat{N}} a_L - q^{\hat{N}} a_R^\dagger \right),
\end{aligned}
\ee
where \( \hat{N} = \hat{n}_L - \hat{n}_R +\Delta\),
denotes the difference of the left and right chord number operators. 
As shown in appendix \ref{sec:commutation}, this identification indeed satisfies the defining commutation relations of the quantum group, as given in \eqref{eq:defining}\footnote{The commutation relations are preserved under linear isomorphisms. In particular, if we define the generators $\mathcal{J} \in \{K, E, F\}$ through a similarity transformation  
\[
\mathcal{J} \;\mapsto\; V_\Delta\, \mathcal{J}\, V_\Delta^{-1},
\]
the same algebraic relations continue to hold. We adopt the present identification between the quantum group generators and the two-sided chord ladder operators because this choice makes the action of these operators on the states defined in~\eqref{eq:statem-def} the most transparent.
}. We verify in Appendix~\ref{app:Rep} that the combinations of chord operators on the right-hand side of ~\eqref{eq:id} act on the states defined in~\eqref{eq:statem-def} to reproduce the discrete representation~\eqref{eq:discrete-rep}. We examine the $q \to 1$ limit of this structure in the following section.

It is important to emphasize that the factor \(\mathcal{L}^2([0,\pi], \mathrm{d}\mu(\phi))\) in \eqref{eq:crossed-product} should not be identified with either the left or the right copy of \(\mathcal{H}_0 \otimes \mathcal{H}_0\). Although a one-sided operator, such as \(a_L\) or \(a_L^{\dagger}\), leaves the right factor of \(\mathcal{H}_0 \simeq \mathcal{L}^2([0,\pi], \mathrm{d}\mu(\phi))\) invariant, it does not act within a fixed representation \(\mathcal{D}^{+}_\Delta\); instead, it maps a state \( |\Delta; m] \) outside of \(\mathcal{D}^{+}_\Delta\), as explicitly shown in~\ref{app:recursion}. The essential feature of the combinations in \eqref{eq:id} is that they preserve the decomposition \eqref{eq:crossed-product}, ensuring that each quantum group generator acts within the \(\mathcal{D}^{+}_\Delta\) factor for any given value of \(\phi\in[0,\pi]\). This structural compatibility is made explicit in Appendix~\ref{app:Rep}.


\subsection{The Coproduct Structure}
As mentioned above, the quantum group $\uqsu$ carries the structure of a Hopf algebra. In particular, this means that it comes equipped with a coproduct. We review the coproduct structure of $\uqsu$ and will utilize it to construct the multi-particle Hilbert space in the following section.

Given an algbera $\mathcal{A}$, a coproduct $\delta: \mathcal{A} \to \mathcal{A} \otimes \mathcal{A}$ is a linear map that satisfies coassociativity:
\be
(\delta \otimes 1) \circ \delta = (1\otimes \delta) \circ  \delta~, 
\ee
with $\circ$ denoting the composition of maps. In addition, it is compatible with the multiplication in the sense that $\delta(ab)=\delta(a)\delta(b)$. The coassociativity of the coproduct ensures that the successive application of the coproduct $\delta$ is well-defined and unambiguous: the position at which the second coproduct acts does not affect the outcome. We therefore introduce the $n$-fold action of the coproduct:
\be
\delta^n : \mathcal{A} \to \mathcal{A}^{\otimes (n+1)}~,
\ee
defined recursively as
\be \label{eq:recursivecoprod}
\delta^n \equiv \delta_{n} \circ \delta_{n-1} \circ \cdots \circ \delta~, \qquad n = 1, 2, \dots,
\ee
where
\be
\delta_{n} = \mathbf{1} \otimes \mathbf{1} \otimes \cdots \otimes \mathbf{1} \otimes \delta :
\mathcal{A}^{\otimes n} \to \mathcal{A}^{\otimes (n+1)}~,
\ee
acts on the $n$-th tensor factor. We shall use this structure to build multi-particle chord Hilbert space in the forthcoming discussion.

To define a coproduct on $\uqsu$, we can specify the action of $\delta$ on generators and extend it using the rule $\delta(ab)=\delta(a)\delta(b)$. The quantum group $\uqsu$ possesses the following coproduct structure:
\be \label{eq:coproduct-QG}
\begin{aligned}
& \delta (K)=K \otimes K, \quad \delta (K^{-1})=K^{-1} \otimes K^{-1}~, \\
& \delta (E)=K \otimes E + E \otimes K^{-1}~, \\
& \delta (F)= K \otimes F+ F \otimes K^{-1}~,
\end{aligned}
\ee 
One can verify the consistency with the aforementioned $*$-structure of $\uqsu$:
\be
\delta(*a) = (*\otimes *)(\delta(a))~, \quad \forall a \in \uqsu~.
\ee
The above coproduct structure can be translated to the chord algebra by defining an action on the chord operators on the right-hand side of \eqref{eq:id}. We find that the coproduct structure \eqref{eq:coproduct-QG} is reproduced by the following coproduct on the chord algebra:\footnote{An alternative coproduct structure was found in the chord algebra in terms of operators acting directly on $\mathcal{H}_1$, see \cite{Lin_2023}. It would be interesting to understand the connection to the construction presented here.}
\be
\begin{aligned}
& \delta (\hat{N})=\hat{N} \otimes 1+1 \otimes \hat{N}~, \\
& \delta (a_L)=a_L \otimes 1+q^{2 \hat{N}} \otimes a_L~, \\
& \delta (a_R)=1 \otimes a_R+a_R \otimes q^{-2 \hat{N}}~.
\end{aligned}
\ee
Moreover, the $*$-structure \eqref{eq:star-structure} is implemented by taking Hermitian conjugates of the chord operators:
\be
*N=N~,\quad*a_{L/R}^{\da}=a_{L/R}~,\quad*a_{L/R}=a_{L/R}^{\da}~.
\ee
We conclude that under these identifications the chord algebra fully encodes the structure of the quantum group $\uqsu$. 

\subsection{Chord Block Decomposition from the Coproduct}

Building on the identification of the one-particle Hilbert space $\mh_1$ with the decomposition $\mathcal{D}^{+}_\Delta \otimes \ml^2([0,\pi],\dd\mu(\phi))$, we now construct multi-particle states in DSSYK. This is achieved by considering the product representation with several copies of $\mh_1$ and subsequently decomposing it into irreducible representations of $\uqsu$, making use of the coproduct structure. Specifically, we will argue that the two-particle Hilbert space $\mh_2^{\Delta_1\Delta_2}$ with two matter chords with weight $\Delta_1$ and $\Delta_2$ is constructed as:
\be \label{eq:2-particle}
\mh_{2}^{\Del_{1}\Del_{2}} \simeq \mathcal{D}_{\Del_{1}}^{+}\otimes \mathcal{D}_{\Del_{2}}^{+}\otimes\mathcal{L}^{2}([0,\pi],\dd\mu(\te_R))~.
\ee
In particular, a two-particle state with fixed boundary energy $|\Delta;\te_L,\phi,\te_R\ket$ can be expressed in terms of states in $\mathcal{D}^{+}_\Delta$ as:
\be \label{eq:2-particleformula}
\begin{split}
    |\Delta_1,\Delta_2;\te_{L},\phi,\te_{R}\ket	=&\left(\sum_{n_{1}=0}^{\infty}\frac{\left(\te_{L},\phi|q^{2\Del_1\hat{n}}\right)Q_{n_{1}}\left(\cos\theta_{L}\mid q^{2\Delta_{1}}e^{\pm i\phi},q^{2}\right)}{\sqrt{\left(q^{4\Delta_{1}},q^{2};q^{2}\right)_{\infty}}}|\Delta_{1};n_{1}]\right) \\
&\otimes\left(\sum_{n_{2}=0}^{\infty}\frac{\left(\phi,\te_{R}|q^{2\Del_2\hat{n}}\right)Q_{n_{2}}\left(\cos\phi\mid q^{2\Delta_{2}}e^{\pm i\te_{R}},q^{2}\right)}{\sqrt{\left(q^{4\Delta_{2}},q^{2};q^{2}\right)_{\infty}}}|\Delta_{2};n_{2}]\right) \otimes|\te_{R})~,
\end{split}
\ee
where we used the labeling $(\te_L,\phi,\te_R)$ to match with the standard notation in DSSYK in later discussion. Note that we construct the two-particle Hilbert space $\mh^{\Delta_1\Delta_2}_2$ in \eqref{eq:2-particle} using representation theory considerations, and it is straightforward to show that the inner product of the right hand side of \eqref{eq:2-particle} agrees with the one given by chord rules. Explicitly, taking the inner product of the states in \eqref{eq:2-particleformula} and using the completeness of Al-Salam Chihara polynomials, we find:
\be
\bra\Del_{1},\Del_{2};\te_{L}^{\pp},\phi^{\pp},\te_{R}^{\pp}|\Del_{1},\Del_{2};\te_{L},\phi,\te_{R}\ket=\gamma_{\Del}^{2}(\te_{L},\phi)\gamma_{\Del}^{2}(\phi^{\pp},\te_{R})\frac{\delta(\te_{L}^{\pp}-\te_{L})\delta(\phi-\phi^{\pp})\del(\te_{R}-\te_{R}^{\pp})}{\mu(\te_{L})\mu(\phi)\mu(\te_{R})}~.
\ee
Therefore, the Hilbert space can be identified with the actual two-particle sector of the chord Hilbert space. 

It is worth pointing out that the same construction has been utilized in \cite{Cui:2025sgy}, motivated by the split representation of amplitudes in sine-dilaton gravity. Here, we motivate our construction \eqref{eq:2-particle} with the coproduct structure of $\uqsu$.  In fact, the state \eqref{eq:2-particleformula} is an eigenstate of the operator \( \delta( Y_{\te_R} K) \) which acts on the first two tensor factors of the state:
\be
\delta(Y_{\te_R}K) = Y_{\te_R }K \otimes \mathbf{1} + K^2 \otimes Y_{\te_R}K~.
\ee
We have
\be \label{eq:eigen-3}
(\delta(Y_{\te_R} K)\otimes \mathbf{1}) |\Delta_1,\Delta_2;\te_L,\phi ,\te_R \ket = E(\te_L,\te_R) |\Delta_1,\Delta_2;\te_L,\phi,\te_R \ket~,
\ee
where $E(\theta_L,\theta_R)$ is given in \eqref{eq:two-sidedenergy}. Note that this is the same eigenvalue as for one-particle states with left and right energy given by $E(\te_{L/R})$. Importantly, it is independent of the intermediate energy label $\phi$. We can now use the spectral theory of the self-adjoint operator $\delta(Y_{\te_R} K)$ to decompose the two-particle Hilbert space $\mh^{\Delta_1\Delta_2}_2$ into one-particle Hilbert spaces. To this end, we introduce the composite matter weight as:\footnote{Throughout, we use $\Delta_j$ with $j$ denoting a non-negative integer, without fixing a specific value. This should not be confused with $\Delta_1$ and $\Delta_2$, which refer to the fixed conformal weights of the two matter chords.}
 \be
\Del_{j}\equiv \Del_{1}+\Del_{2}+j~,\qquad j=0,1,2,\dots,
 \ee
The eigenvalue equation \eqref{eq:eigen-3} implies the existence of an intertwiner \cite{koelink1996}:
\be
\Upsilon:\mh_{2}^{\Del_{1}\Del_{2}}\to\bigoplus_{j=0}^{\infty}\mh_{1}^{\Del_{j}}~,
\ee
Note that $\Upsilon$ is not to be confused with the intertwiner $\Upsilon_\Delta$ we defined in previous section. The above map intertwines the action of $\delta(Y_{\te_R}K)$ with the action of $Y_{\te_R} K$ on the one-particle Hilbert spaces:
\be
\Upsilon\circ\delta(Y_{\te_{R}}K)=Y_{\te_{R}}K\circ\Upsilon~. 
\ee
Explicitly, we have the following identification:
\be \label{eq:intertwine-2p}
\Upsilon|\Delta_{1},\Delta_{2};\te_{L},\phi,\te_{R}\ket=\sum_{j=0}^{\infty} \frac{\gamma_{\Del_{1}}(\te_{L},\phi)\gamma_{\Del_{2}}(\phi,\te_{R})}{\gamma_{\Del_{j}}(\te_{L},\te_{R})} P_{j}^{\Del_{1}\Del_{2}}(\cos\phi)|\Del_{j};\te_{L},\te_{R}\ket~,
\ee
where the coefficient $P^{\Delta_1\Delta_2}_j (\cos\phi)$ is a $3j$-symbol of the quantum group $\uqsu$. It is a polynomial of degree $j$ in its argument $\cos\phi$, and defined explicitly in terms of the so-called Askey-Wilson polynomial $p_j(x;q^a,q^b,q^c,q^d;q)$ \cite{jafferis2023} as:
\be
P_{j}^{\Del_{1}\Del_{2}}(\cos\phi)\equiv \frac{\gamma_{\Del_{1}}(\te_{L},\phi)\gamma_{\Del_{2}}(\phi,\te_{R})}{\gamma_{\Del_{j}}(\te_{L},\te_{R})}\frac{p_{j}\left(\cos\phi;q^{2\Del_{1}}e^{\pm i\te_{L}},q^{2\Del_{2}}e^{\pm i\te_{R}};q^{2}\right)}{\sqrt{\left(q^{4\Del_{1}},q^{4\Del_{2}},q^{4\Del_{j}-2},q^{2};q^{2}\right)_{j}}}~.
\ee
Here the $\pm$ symbol indicates that the arguments of the function are filled with alternating phases.  
Specifically, this corresponds to the assignment  
\be\label{eq:assignment}
(q^{a}, q^{b}, q^{c}, q^{d}) = 
\big(q^{2\Delta_{1}} e^{i\theta_{L}},\,
q^{2\Delta_{1}} e^{-i\theta_{L}},\,
q^{2\Delta_{2}} e^{i\theta_{R}},\,
q^{2\Delta_{2}} e^{-i\theta_{R}}\big)~.
\ee 
The unitarity of $\Upsilon$ follows from the orthogonality and completeness conditions of the $3j$-symbols\footnote{For a derivation, see proposition 4.3 in~\cite{koelink1996}.}:
\be \label{eq:3jcompleteness}
\begin{split}
&\int_{0}^{\pi}\dd\mu(\phi)\,P_{i}^{\Del_{1}\Del_{2}}(\cos\phi)P_{j}^{\Del_{1}\Del_{2}}(\cos\phi)=\delta_{ij}~,\\  &\sum_{j=0}^{\infty}P_{i}^{\Del_{1}\Del_{2}}(\cos\phi)P_{j}^{\Del_{1}\Del_{2}}(\cos\phi)=\frac{\delta(\phi-\phi^{\pp})}{\mu(\phi)}~.
\end{split}
\ee
We introduce the adjoint 
\be \label{eq:intertwine-2pd}
\Upsilon^{\da}|\Del_{j};\te_{L},\te_{R}\ket=\int_{0}^{\pi}\dd\mu(\phi) \frac{\gamma_{\Del_{j}}(\te_{L},\te_{R})}{\gamma_{\Del_{1}}(\te_{L},\phi)\gamma_{\Del_{2}}(\phi,\te_{R}) }  P_{j}^{\Del_{1}\Del_{2}}(\cos\phi)|\Delta_{1},\Delta_{2};\te_{L},\phi,\te_{R}\ket~,
\ee 
where the coefficients $\gamma_{\Delta}$ in \eqref{eq:intertwine-2p} and \eqref{eq:intertwine-2pd} reflect the normalization of states on the two sides. They are normalized to agree with the corresponding fixed energy correlator in DSSYK. Using the identities in \eqref{eq:3jcompleteness} it is easy to check that:
\be
\Upsilon\Upsilon^{\da}=\mathbf{1}_{\oplus_{j=0}^{\infty}\mh_{1}^{\Del_{j}}}~,\quad\Upsilon^{\da}\Upsilon=\mathbf{1}_{\mh_{2}^{\Del_{1}\Del_{2}}}~.
\ee
\paragraph{Implementing Crossing via the Quantum $6j$-Symbol} 
We now examine the action of the $R$-matrix and determine the extent to which it implements the crossing of two matter chords in the bulk.
To this end, we introduce an operator $\mathcal{R}$ acting on the two-particle Hilbert space $\mathcal{H}^{\Delta_1\Delta_2}_2$, defined by
\be \label{eq:R-matrix-def}
\begin{split}
\mathcal{R}\,|\Delta_{1},& \Delta_{2}; \theta_{L},\phi,\theta_{R}\rangle  \\
&\equiv  \int_{0}^{\pi}\dd\mu(\phi') \,
\frac{\gamma_{\Delta_{1}}(\theta_{R},\phi)\,\gamma_{\Delta_{2}}(\theta_{L},\phi)}
{\gamma_{\Delta_{1}}(\theta_{L},\phi')\,\gamma_{\Delta_{2}}(\theta_{R},\phi')}
q^{2\Delta_1\Delta_2}\left\{ 
\begin{array}{lll}
\Delta_{1} & \theta_{L} & \phi' \\
\Delta_{2} & \theta_{R} & \phi
\end{array}
\right\}_{q^2}
|\Delta_{1},\Delta_{2};\theta_{L},\phi',\theta_{R}\rangle~,
\end{split}
\ee
where the integration kernel is the $\uqsu$ quantum $6j$-symbol associated with a crossed four-point function.  
For an explicit expression of the $6j$-symbol in terms of special functions, we refer the reader to the original derivation in~\cite{Berkooz:2018jqr}, which was obtained from chord combinatorics.   Based on these results, the action of $\mathcal{R}$ can thus be understood as implementing a swap of matter chords \cite{Berkooz:2024lgq}:
\be \label{eq:swap}
\begin{tikzpicture}[baseline={([yshift=-0.1cm]current bounding box.center)},scale=0.7,thick]

\draw[black] (-2,0) arc[start angle=-180, end angle=0, radius=2];

\draw[blue, thick] 
    (1,-1.73) 
    .. controls (0.7,-1.4) and (-0.7,-0.6) ..
    (-1,0);

\draw[red, thick] 
    (-1,-1.73) 
    .. controls (-0.7,-1.4) and (0.7,-0.6) ..
    (1,0);

\node at (-2.2,-1.2) {$\theta_L$};
\node at (2.2,-1.2) {$\theta_R$};
\node at (0,-2.4) {$\phi$};

\node[blue] at (-1,0.2) {$\Delta_2$};
\node[red] at (1,0.2) {$\Delta_1$};

\end{tikzpicture}
=
\int^{\pi}_0 \dd\mu(\phi^\pp) \mathcal{R}_{\phi\phi^\pp}\begin{bmatrix}\Del_{1} & \te_L\\
\Del_{2} & \te_R
\end{bmatrix}
\begin{tikzpicture}[baseline={([yshift=-0.1cm]current bounding box.center)},scale=0.7,thick]

\draw[black] (-2,0) arc[start angle=-180, end angle=0, radius=2];

\draw[blue, thick] 
    (1,-1.73) -- (1,0);

\draw[red, thick] 
    (-1,-1.73)  -- (-1,0);

\node at (-2.2,-1.2) {$\theta_L$};
\node at (2.2,-1.2) {$\theta_R$};
\node at (0,-2.4) {$\phi^\pp$};

\node[blue] at (1,0.2) {$\Delta_2$};
\node[red] at (-1,0.2) {$\Delta_1$};

\end{tikzpicture}~,
\ee
where we have denoted the components in \eqref{eq:R-matrix-def} as the $R$-matrix in \eqref{eq:swap}.  Here, we present a group-theoretical perspective: using the intertwiner $\Upsilon$, we decompose a given two-particle state into one-particle channels and analyze the action of the operator $\mathcal{R}$ within each channel.

Using \eqref{eq:intertwine-2p}, we find that
\be \label{eq:R-action}
\begin{aligned}
\Upsilon \circ \mathcal{R}\,|\Delta_1, \Delta_2 & ; \theta_L, \phi, \theta_R\rangle 
= \gamma_{\Delta_1}(\theta_R, \phi)\,\gamma_{\Delta_2}(\theta_L, \phi) \\
& \quad \times \sum_{j=0}^{\infty} 
(-1)^j \, q^{\,2[\Delta_1\Delta_2+j(\Delta_1+\Delta_2) + \frac{j(j-1)}{2}]} \,
\frac{P_j^{\Delta_2 \Delta_1}(\cos \phi)}{\gamma_{\Delta_j}(\theta_L, \theta_R)}
\,|\Delta_j ; \theta_L, \theta_R\rangle~,
\end{aligned}
\ee
where we have made use of the following Askey--Wilson transformation \cite{jafferis2023}: 
\be
\int_{0}^{\pi}\dd\mu(\phi') 
\left\{ 
\begin{array}{lll}
\Delta_{1} & \te_L & \phi' \\
\Delta_{2} & \te_R & \phi
\end{array}
\right\}_{q^2}
P_{n}^{\Delta_{1}\Delta_{2}}(\cos\phi')
= (-1)^{n} q^{2[\,n(\Delta_{1}+\Delta_{2})+\frac{n(n-1)}{2}]}
P_{n}^{\Delta_{2}\Delta_{1}}(\cos\phi)~.
\ee
Here, we have kept the standard phase convention for $3j$-symbols, so that a factor of $(-1)^n$ appears in each intermediate channel. 
The additional composite penalty factor 
$q^{2[\,n(\Delta_1+\Delta_2)+n(n-1)/2]}$ 
can be understood combinatorially: In each one-particle fusion channel of a matter chord $\textcolor{red}{V}$ with weight $\Delta_1$ and a matter chord $\textcolor{blue}{W}$ with weight $\Delta_2$, there exists a composite chord of the form $\textcolor{red}{V} H^{n} \textcolor{blue}{W}$.  
The action of $\mathcal{R}$ reverses the ordering of the chords, transforming $\textcolor{red}{V} H^{n} \textcolor{blue}{W}$ into $\textcolor{blue}{W} H^{n} \textcolor{red}{V}$, while the composite penalty factor accounts for the number of crossings generated during this process.  
This procedure can be illustrated schematically as follows:
\be
\begin{tikzpicture}[baseline={([yshift=0.15cm]current bounding box.center)},scale=0.4,thick]

\draw[red, thick] 
    (-2,-1)  -- (-2,1);
\draw[blue, thick] 
    (1,-1) -- (1,1);
\draw[thick] (-1.5,-1)--(-1.5,1);
\draw[thick] (-1,-1)--(-1,1);

\node at (-2.4,-1.6) {\small $\textcolor{red}{V}$};
\node at (1,-1.6) {\small $\textcolor{blue}{W}$};
\node at (0,0) {$\cdots$};
\end{tikzpicture}
\longrightarrow \hspace{8pt}
\begin{tikzpicture}[baseline={([yshift=-0.05cm]current bounding box.center)},scale=0.4,thick]

\draw[red, thick] 
    (-2,-1)  -- (-2,-0.2) -- (1.5,-0.2)--(1.5,1);
\draw[blue, thick] 
    (1,-1) -- (1,1);
\draw[thick] (-1.5,-1)--(-1.5,1);
\draw[thick] (-1,-1)--(-1,1);

\node at (0,0) {$\cdots$};
\end{tikzpicture}
\hspace{8pt}\longrightarrow\hspace{8pt}
\begin{tikzpicture}[baseline={([yshift=-0.05cm]current bounding box.center)},scale=0.4,thick]

\draw[red, thick] 
    (2.0,-1)--(2.0,1);
\draw[blue, thick] 
    (1,-1) -- (1,1);
\draw[thick] 
(-1.5,-1)--(-1.5,-0.2)--(1.5,-0.2)--(1.5,1);
\draw[thick] (-1,-1)--(-1,1);

\node at (0,0) {$\cdots$};
\end{tikzpicture}
\hspace{8pt}\longrightarrow \hspace{8pt} \cdots 
\hspace{8pt}\longrightarrow
\begin{tikzpicture}[baseline={([yshift=0.15cm]current bounding box.center)},scale=0.4,thick]

\draw[blue, thick] 
    (-2,-1)  -- (-2,1);
\draw[red, thick] 
    (1,-1) -- (1,1);
\draw[thick] (-1.5,-1)--(-1.5,1);
\draw[thick] (-1,-1)--(-1,1);

\node at (-2.4,-1.6) {\small $\textcolor{blue}{W}$};
\node at (1,-1.6) {\small $\textcolor{red}{V}$};
\node at (0,0) {$\cdots$};
\end{tikzpicture}
\ee
The composite penalty factor of this move is therefore
\be
\begin{split}
    \Delta_{\textcolor{red}{1}}\,(n+\Delta_{\textcolor{blue}{2}})  + (n-1+\Delta_{\textcolor{blue}{2}}) &+ (n-2+\Delta_{\textcolor{blue}{2}}) + \cdots + \Delta_{\textcolor{blue}{2}}\\ 
    &=\Delta_{\textcolor{red}{1}}\Delta_{\textcolor{blue}{2}}+ (\Delta_{\textcolor{red}{1}}+\Delta_{\textcolor{blue}{2}})\,n + \tfrac{n(n-1)}{2}~.
\end{split}
\ee
Taking inner product of \eqref{eq:R-action} with the image of a two-particle state under $\Upsilon$, and utilizing the unitarity of the intertwiner, we find that
\be
\begin{split}
\bra\Delta_{1},\Delta_{2};\te_{L}^{\pp},\phi^{\pp},\te_{R}^{\pp}|\mathcal{R}|\Delta_{1},\Delta_{2};\te_{L},\phi,\te_{R}\ket	=&\gamma_{\Del_{1}}(\te_{R},\phi)\gamma_{\Del_{1}}(\te_{L},\phi^{\pp})\gamma_{\Del_{2}}(\te_{L},\phi)\gamma_{\Del_{2}}(\phi^{\pp},\te_{R}) \\
	& \times q^{2\Delta_1\Delta_2}\left\{ \begin{array}{lll}
\Delta_{1} & \te_{L} & \phi^{\pp}\\
\Delta_{2} & \te_{R} & \phi
\end{array}\right\} _{q^2},
\end{split}
\ee
which equals the crossed four-point function with fixed energies $(\theta_L,\phi',\theta_R,\phi)$ in DSSYK.

\subsection{Construction of Multi-Particle Hilbert Space from the Coproduct}

In this section, we construct the multi-particle chord Hilbert space. To begin, it is convenient to introduce a state that represents an eigenstate of $Y_\phi K$:
\be
|\Delta;\theta]_{\phi} \equiv (\theta,\phi|q^{2\Delta \hat{n}}) 
\sum_{n=0}^{\infty} 
\frac{Q_{n}(\cos\theta|q^{2\Delta}e^{\pm i\phi},q^{2})}
{\sqrt{(q^{2},q^{4\Delta};q^{2})_{n}}} \, |\Delta;n]~,
\ee
so that the one-particle state \eqref{eq:eigenstate-2} can be written compactly as
\be
|\Delta;\theta_{L},\theta_{R}\ket = |\Delta;\theta_{L}]_{\theta_{R}} \otimes |\theta_{R})~,
\ee
where the lower label $\te_R$ makes the $\te_R$ dependence of the state explicit.  We now define the $n$-particle Hilbert space $\mathcal{H}_n(\Delta_1,\dots,\Delta_n)$, spanned by states of the form
\be \label{eq:decompose-n}
|\Delta_{1},\Delta_{2},\dots,\Delta_{n};\phi_{0},\phi_{1},\dots,\phi_{n-1},\phi_{n}\ket 
\equiv |\Delta_{1};\phi_{0}]_{\phi_{1}} \otimes |\Delta_{2};\phi_{1}]_{\phi_{2}} 
\otimes \cdots \otimes |\Delta_{n};\phi_{n-1}]_{\phi_{n}} \otimes |\phi_{n})~,
\ee
with each $\phi_i \in [0,\pi], \, i=0,1,\dots,n$ distributed according to the $0$-particle energy density $\dd\mu(\phi_i)$.  
This decomposition was previously obtained via a split representation in sine–dilaton gravity \cite{Cui:2025sgy}. Here, the state \eqref{eq:decompose-n} is constructed as an eigenstate of $\delta^{n-1}(Y_{\phi_n} K)$ with eigenvalue $E(\phi_0,\phi_n)$, generalizing our earlier two-particle construction. Note that this construction essentially follows from the coproduct structure of $\uqsu$, thereby providing a microscopic origin for the split representation that, in the bulk description, arises as a property of the gravitational amplitude~\cite{Cui:2025sgy}.  
It would be interesting to further elucidate how this algebraic structure manifests in the dual gravitational description.

Denoting $|\Psi_n\ket = |\Delta_{1},\dots,\Delta_{n};\phi_{0},\dots,\phi_{n}\ket$, we have the following eigenvalue equation:
\be \label{eq:n-eigenstate-eqn}
\delta^{n-1}(Y_{\phi_{n}}K) |\Psi_n\ket = E(\phi_{0},\phi_{n}) |\Psi_n \ket~,
\ee
which we will prove at the end of this section.

Applying \eqref{eq:reverse} to \eqref{eq:decompose-n}, we find that the fixed chord number states can be expressed as:
\be
\Bigl(\otimes_{k=1}^{n} |\Delta_{k},n_{k-1}]\Bigr) \otimes |\phi_{n}) 
= \int \prod_{k=1}^{n} \dd\mu(\phi_{k-1}) 
\frac{Q_{n_{k-1}}\!\left(\cos\phi_{k-1}\middle|q^{2\Delta}e^{\pm i\phi_{k}};q^{2}\right)}
{\sqrt{(q^{4\Delta},q^{2};q^{2})_{n_{k-1}}}} \, |\Psi_n \ket~,
\ee
which makes manifest how the Hilbert space $\mathcal{H}_n(\Delta_1,\dots,\Delta_n)$ decomposes into a tensor product of positive discrete series representations of the quantum group:
\be
\mathcal{H}_{n}(\Delta_{1},\dots,\Delta_{n})
\simeq \bigotimes_{k=1}^{n} \mathcal{D}^{+}_{\Delta_{k}} 
\otimes \mathcal{L}^{2}\!\left([0,\pi];\dd\mu(\phi_{n})\right)~.
\ee

We now prove the eigenvalue equation \eqref{eq:n-eigenstate-eqn}.  
Recalling the definition \eqref{eq:recursivecoprod}, the successive action of the coproduct yields
\be \label{eq:delta-n-1}
\delta^{n-1}(Y_{\phi_{n}}K) =
Y_{\phi_{n}}K \otimes \mathbf{1}
+ K^{2} \otimes Y_{\phi_{n}}K \otimes \mathbf{1}
+ \cdots
+ K^{2} \otimes \cdots \otimes K^{2} \otimes Y_{\phi_{n}}K~.
\ee
Denoting by $K_j$ the operator $K$ acting on the $j$-th tensor factor, and by $(Y_\phi K)_l$ the operator $Y_\phi K$ acting on the $l$-th factor, we can rewrite \eqref{eq:delta-n-1} as
\be \label{eq:delta-n-2}
\delta^{n-1}(Y_{\phi_{n}}K)
= \sum_{l=1}^{n}
\left( \prod_{j=1}^{l-1} K_{j}^{2} \right) (Y_{\phi_{n}}K)_{l}~.
\ee
Consider $(Y_{\phi_n}K)_1$ acting on the first copy of $\mathcal{D}^{+}_{\Delta_1}$.  
Using \eqref{eq:YK-def}, we can replace $Y_{\phi_n}K$ by $Y_{\phi_1}K$ up to a correction proportional to $(K_{1}^{2}-1)$:
\be
(Y_{\phi_{n}}K)_{1}
= (Y_{\phi_{1}}K)_{1} - E(\phi_{1},\phi_{n})(K_{1}^{2}-1)~.
\ee
From the decomposition \eqref{eq:decompose-n}, $Y_{\phi_1}K$ acts diagonally on the first factor $|\Delta_1;\phi_0]_{\phi_1}$ with eigenvalue $E(\phi_0,\phi_1)$.  
Thus, the action of $(Y_{\phi_n}K)_1$, i.e. the first term in \eqref{eq:delta-n-2}, gives
\be
(Y_{\phi_{n}}K)_{1} |\Psi_{n}\ket
= \left[ E(\phi_{0},\phi_{1}) - E(\phi_{1},\phi_{n})(K_{1}^{2}-1) \right] |\Psi_{n}\ket~.
\ee
Similarly, replacing $(Y_{\phi_n}K)_l$ with $(Y_{\phi_l}K)_l$ for general $l$ gives
\be
\left(\prod_{j=1}^{l-1} K_{j}^{2}\right) (Y_{\phi_{n}}K)_{l} |\Psi_{n}\ket
= \left[ E(\phi_{l-1},\phi_{l}) \prod_{j=1}^{l-1} K_{j}^{2}
- E(\phi_{l},\phi_{n}) \left( \prod_{j=1}^{l} K_{j}^{2} - \prod_{j=1}^{l-1} K_{j}^{2} \right) \right] |\Psi_{n}\ket~.
\ee
Combining this with \eqref{eq:delta-n-2}, we obtain
\be
\delta^{n-1}(Y_{\phi_{n}}K) |\Psi_{n}\ket
= \sum_{l=1}^{n}
\left[ E(\phi_{l-1},\phi_{l}) \prod_{j=1}^{l-1} K_{j}^{2}
- E(\phi_{l},\phi_{n}) \left( \prod_{j=1}^{l} K_{j}^{2}
- \prod_{j=1}^{l-1} K_{j}^{2} \right) \right] |\Psi_{n}\ket~.
\ee
The $l=1$ term in the sum produces the desired eigenvalue relation: 
\be
E(\phi_{0},\phi_{1}) + E(\phi_{1},\phi_{n}) = E(\phi_{0},\phi_{n})~,
\ee
while all terms with $l \ge 2$ cancel among themselves:
\be
\sum_{l=2}^{n}
\left[ E(\phi_{l-1},\phi_{l}) - E(\phi_{l-1},\phi_{n}) + E(\phi_{l},\phi_{n}) \right]
\prod_{j=1}^{l} K_{j}^{2} |\Psi_{n}\ket = 0~,
\ee
where we have used \eqref{eq:E2-def} to simplify
\be
E(\phi_{l-1},\phi_l) - E(\phi_{l-1},\phi_n)
= E(\phi_n,\phi_l)
= -E(\phi_l,\phi_n)~.
\ee
This confirms \eqref{eq:n-eigenstate-eqn}.  

The spectral decomposition of $\delta^{n-1}(Y_{\phi_n}K)$ now allows the construction of an intertwiner analogous to the two-particle case \eqref{eq:intertwine-2p}, enabling the decomposition of $\mathcal{H}_n$ into a direct sum of one-particle states with distinct weights.  
In practice, one can successively fuse the first two particles into a composite one, then fuse it with the third, and so on, until all particles are fused into a single composite one-particle state.  
The isometry of the intertwiner guarantees that all information about the chord inner product, originally defined combinatorially by chord rules, is faithfully encoded in the fusion rules, which are fully determined by the quantum group structure.

\subsection{Deriving the Yang-Baxter Relation}
It was observed in \cite{Lin_2023} that chord diagrams in the double-scaled SYK model exhibit a Yang–Baxter–type invariance: they remain unchanged under specific re-orderings of mutually intersecting chords. This invariance reflects the fact that the chord combinatorics depend only on the number of bulk intersections, not on the particular order in which those intersections are realized.

To understand the extent to which the chord combinatorial rules give rise to a notion of bulk geometry, it is essential to uncover the microscopic mechanism that realizes the chord Yang–Baxter relation. In this section, we derive the this relation directly from the chord block decomposition established in the previous section, thereby providing a concrete representation theoretic origin for this relation.

We begin with the following identity \cite{stokman2001} involving the Askey-Wilson polynomials and quantum $6j$-symbols:
\be \label{eq:identity-1}
\begin{aligned}
\int_{0}^{\pi} \dd \mu(\theta_{f}) 
\left[ q^{2\Delta_{1} \Delta_{3}}  \left\{ \begin{array}{ccc}
\Delta_{1} & \theta_{b} & \theta_{c} \\
\Delta_{3} & \theta_{f} & \theta_{a}
\end{array} \right\}_{q^2} \right] &
\left[ q^{2\Delta_{2} \Delta_{3}} \left\{ \begin{array}{ccc}
\Delta_{2} & \theta_{c} & \theta_{d} \\
\Delta_{3} & \theta_{e} & \theta_{f}
\end{array} \right\}_{q^2} \right]
P_{n}^{\Delta_{1} \Delta_{2}}(\cos \theta_{f})\\
= q^{2(\Delta_{1}+\Delta_{2}+n)\Delta_{3}}& \left\{ \begin{array}{ccc}
\Delta_{1}+\Delta_{2}+n & \theta_{b} & \theta_{d} \\
\Delta_3 & \theta_{e} & \theta_{a}
\end{array} \right\}_{q^2} P_{n}^{\Delta_{1} \Delta_{2}}(\cos \theta_{c}) ~,
\end{aligned}
\ee
which originates from the pentagon identity satisfied by the $6j$-symbol.\footnote{For a derivation, we refer readers to Appendix G of \cite{jafferis2023}}.

In the chord diagram formulation, each crossing of matter chords corresponds to a quantum $6j$-symbol in the energy eigenbasis, while each Askey–Wilson polynomial represents the fusion vertex of two matter chords into a composite one.  In terms of our construction of multi-particle Hilbert space and fusion rules in previous section, \eqref{eq:identity-1} can be expressed simply as:
\be \label{eq:simple-rep}
 \bra \Delta_1,\Delta_2;\te_b,\te_c,\te_d|q^{2\Delta_3\hat{N}^{(2)}_{tot}}\circ \Upsilon^\dagger|\Delta_n;\te_a,\te_e\ket = \bra \Delta_1,\Delta_2;\te_b,\te_c,\te_d|\Upsilon^\dagger\circ q^{2\Delta_3\hat{N}^{(1)}_{tot}}|\Delta_n;\te_a,\te_e\ket~.
\ee
Here, $\hat{N}^{(2)}_{\text{tot}}$ and $\hat{N}^{(1)}_{\text{tot}}$ denote the total chord number operators in the two-particle and one-particle sectors, respectively.  
Their matrix elements were computed explicitly in~\cite{Cui:2025sgy}, and with those results, the representation~\eqref{eq:simple-rep} is seen to agree precisely with~\eqref{eq:identity-1}.
The above equation can be illustrated as follows:
\be \label{eq:identity-2}
\begin{tikzpicture}[scale=2.2, baseline={(current bounding box.center)}]
  \coordinate (A1) at (-2,1);
  \coordinate (A2) at (-1,1);
  \coordinate (A3) at (-1.5,0.5);
  \coordinate (A4) at (-1.5,-0.25);
  \coordinate (AH) at (-2.3,0.75);
  \coordinate (AH2) at (-0.7,0.75);
  \draw[thick] (A1) -- (A3);
  \draw[thick] (A2) -- (A3);
  \draw[thick] (A3) -- (A4);
  \draw[thick] (AH) -- (AH2);
  \node at (-2.1,1.15) {$\Delta_1$};
  \node at (-0.9,1.15) {$\Delta_2$};
  \node at (-1.5,-0.5) {$\Delta_1+\Delta_2+n$};
  \node at (-0.8,.9) {$\theta_d$};
  \node at (-2.2, .9) {$\theta_b$};
  \node at (-0.8,0) {$\theta_e$}; 
  \node at (-2.2,0) {$\theta_a$};
  \node at (-1.5,1) {$\theta_c$};
  \node at (-1.5,0.65) {$\theta_f$};
  \node at (0,0.5) {$=$};
  \coordinate (B1) at (1,1);
  \coordinate (B2) at (2,1);
  \coordinate (B3) at (1.5,0.5);
  \coordinate (B4) at (1.5,-0.25);
  \coordinate (BH) at (0.7,0.25);
  \coordinate (BH2) at (2.3,0.25);
  \draw[thick] (B1) -- (B3);
  \draw[thick] (B2) -- (B3);
  \draw[thick] (B3) -- (B4);
  \draw[thick] (BH) -- (BH2);
  \node at (0.9,1.15) {$\Delta_1$};
  \node at (2.1,1.15) {$\Delta_2$};
  \node at (1.5,-0.5) {$\Delta_1+\Delta_2+ n$};
  \node at (0.8,.65) {$\theta_b$};
  \node at (2.2, .65) {$\theta_d$};
  \node at (0.8,0) {$\theta_a$}; 
  \node at (2.2,0) {$\theta_e$};
  \node at (1.5,1) {$\theta_c$};
\end{tikzpicture}
\ee

The diagrammatic identity illustrates the structural equivalence of two procedures: On the left-hand side, a third matter chord is inserted before the fusion of \( \Delta_1 \) and \( \Delta_2 \). This insertion introduces a closed loop in the chord diagram, disconnected from the boundary, requiring a sum over intermediate energy states \( \theta_f \). This resembles a loop integral of Feynman diagrams in standard perturbative quantum field theory. On the right-hand side, the same matter chord is inserted after fusing \( \Delta_1 \) and \( \Delta_2 \), leading to a tree diagram involving a crossing of matter lines and hence a multiplication by a quantum $6j$-symbol.

The prefactors such as \( q^{2\Delta_1 \Delta_3} \) are associated with crossings of matter chords and can be interpreted as interaction ``penalty" weights due to their intersection in the bulk. Importantly, this identity demonstrates that, for generic matter weights, the insertion of a matter chord commutes with the fusion of other matter chords—an expression of the associativity encoded by the pentagon identity.

Since any loop configuration in a chord diagram can be recursively reduced to a collection of tree-level diagrams by successive applications of this identity, the structure of arbitrary matter configurations in the energy basis admits a complete decomposition into tree-level building blocks. Each application of the identity performs a fusion and reorganization of the matter chords, replacing sums over intermediate energy states with compositions of fusion vertices and $6j$-symbols. This recursive reduction yields a systematic method for evaluating chord diagrams with fixed energy boundary conditions.

In \cite{Lin_2023}, it was observed that the chord rules satisfy a Yang-Baxter-type relation, represented diagrammatically as:
\be
\begin{tikzpicture}[scale=1.2, baseline={(current bounding box.center)}, every node/.style={font=\small}]
\coordinate (A1) at (-1,1);
\coordinate (A2) at (1,-1);
\coordinate (B1) at (-1,-1);
\coordinate (B2) at (1,1);
\coordinate (C1) at (-1.2,0.5);
\coordinate (C2) at (1.2,0.5); 

\draw[thick] (A1) -- (A2);
\draw[thick] (B1) -- (B2);
\draw[thick] (C1) -- (C2);

\node at (-1,1.2) {$\Delta_1$};
\node at (1,1.2) {$\Delta_2$};
\node at (-1.5,0.5) {$\Delta_3$};

\coordinate (A5) at (3,1);
\coordinate (A6) at (5,-1);
\coordinate (B5) at (3,-1);
\coordinate (B6) at (5,1);
\coordinate (C5) at (2.8,-0.5);
\coordinate (C6) at (5.2,-0.5);

\draw[thick] (A5) -- (A6);
\draw[thick] (B5) -- (B6);
\draw[thick] (C5) -- (C6);

\node at (3,1.2) {$\Delta_1$};
\node at (5,1.2) {$\Delta_2$};
\node at (5.5,-0.5) {$\Delta_3$};

\node at (2,0) {$=$};
\end{tikzpicture}
\ee

This relation implies that the putative bulk geometry defined by chord diagrams should be understood only up to equivalence classes determined by such identities. While this invariance is manifest at the level of combinatorial chord rules, it is significantly more subtle to derive from the inner product among states with fixed boundary energy, which relies on slicing open diagrams and summing over all possible intermediate chord configurations. Such a derivation typically requires intricate accounting of the $q$-deformed combinatorics associated with chord overlaps.

However, based on the quantum group structure we have established, we can now derive this Yang-Baxter relation using the fusion rules of $\uqsu$. In particular, this relation emerges from the associativity of the fusion of positive discrete series representations. Diagrammatically, the Yang-Baxter move corresponds to two different fusion orderings of three incoming matter chords, both leading to the same final configuration. This equivalence is rooted in the pentagon identity and the representation theory of the underlying quantum group. We depict the fusion-based process as follows:

\be
\begin{tikzpicture}[scale=1.2, baseline={(current bounding box.center)}, every node/.style={font=\small}]
  \coordinate (A11) at (-1,1);
  \coordinate (A21) at (1,-1);
  \coordinate (B11) at (-1,-1);
  \coordinate (B21) at (1,1);
  \coordinate (C11) at (-1.2,0.5);
  \coordinate (C21) at (1.2,0.5);
\node at (-1,1.2) {$\Delta_1$};
\node at (1,1.2) {$\Delta_2$};
\node at (-1.5,0.5) {$\Delta_3$};

  \coordinate (A12) at (3,1);
  \coordinate (A22) at (5,1);
  \coordinate (A32) at (4,0.5);
  \coordinate (B12) at (3,-1);
  \coordinate (B22) at (5,-1);
  \coordinate (B32) at (4,-0.5);
  \coordinate (C12) at (3,.75);
  \coordinate (C22) at (5,.75); 
  \coordinate (A13) at (7,1);
  \coordinate (A23) at (9,1);
  \coordinate (A33) at (8,0.5);
  \coordinate (B13) at (7,-1);
  \coordinate (B23) at (9,-1);
  \coordinate (B33) at (8,-0.5);
  \coordinate (C13) at (7,0);
  \coordinate (C23) at (9,0); 
  \coordinate (A14) at (3,-2);
  \coordinate (A24) at (5,-2);
  \coordinate (A34) at (4,-2.5);
  \coordinate (B14) at (3,-4);
  \coordinate (B24) at (5,-4);
  \coordinate (B34) at (4,-3.5);
  \coordinate (C14) at (3,-3.75);
  \coordinate (C24) at (5,-3.75); 
  \coordinate (A15) at (7,-2);
  \coordinate (A25) at (9,-4);
  \coordinate (B15) at (7,-4);
  \coordinate (B25) at (9,-2);
  \coordinate (C15) at (6.8,-3.5);
  \coordinate (C25) at (9.2,-3.5);

  \draw[thick] (A11) -- (A21);
  \draw[thick] (B11) -- (B21);
  \draw[thick] (C11) -- (C21);
  
  \draw[thick] (A12) -- (A32);
  \draw[thick] (A22) -- (A32);
  \draw[thick] (B12) -- (B32);
  \draw[thick] (B22) -- (B32);
  \draw[thick] (A32) -- (B32);
  \draw[thick] (C12) -- (C22);

  \draw[thick] (A13) -- (A33);
  \draw[thick] (A23) -- (A33);
  \draw[thick] (B13) -- (B33);
  \draw[thick] (B23) -- (B33);
  \draw[thick] (A33) -- (B33);
  \draw[thick] (C13) -- (C23);

  \draw[thick] (A14) -- (A34);
  \draw[thick] (A24) -- (A34);
  \draw[thick] (B14) -- (B34);
  \draw[thick] (B24) -- (B34);
  \draw[thick] (A34) -- (B34);
  \draw[thick] (C14) -- (C24);

  \draw[thick] (A15) -- (A25);
  \draw[thick] (B15) -- (B25);
  \draw[thick] (C15) -- (C25);

  \node at (2,0) {$=\sum_n (-1)^n $};
  \node at (2,-3) {$=\sum_n (-1)^n $};
  \node at (4.2, -0.25) {$n$};
  \node at (4.2, -3.25) {$n$};
  \node at (6,-3) {$=$};
  \node at (6,0) {$=\sum_n (-1)^n $};
  \node at (8.2, -0.25) {$n$};
  \node at (6,0) {$=\sum_n (-1)^n $};
\end{tikzpicture}
\ee 
where in the first step, we perform the fusion of $\Delta_1$ and $\Delta_2$, which involves summing over all intermediate one-particle states in the corresponding channel. We then apply the identities \eqref{eq:identity-1} and \eqref{eq:identity-2} to move the intersecting chord of weight $\Delta_3$ from the top of the diagram to the bottom. Finally, we use the inverse of this identity to rewrite the resulting sum over fusion channels as a single crossing between $\Delta_1$ and $\Delta_2$, with $\Delta_3$ now intersecting from below.

It is also illuminating to reformulate the above discussion in an operator language, building on the structures we have established.  
Since the action of $\mathcal{R}$ on each one-particle channel amounts to multiplication by a constant~\eqref{eq:R-action}\footnote{More precisely, it depends on the parameters $\Delta_1$, $\Delta_2$, and on the label $j$ corresponding to the fusion channel $\Delta_j$.  
The dependence on $\Delta_1$ and $\Delta_2$ arises from the definition of the operator $\mathcal{R}$ itself, while both $\Delta_{1,2}$ and $j$ remain fixed for states belonging to the subspace $\mathcal{H}^{\Delta_j}_1$.
}, it commutes with any operator acting within the one-particle sector.  
Defining 
\be
\tilde{R} \equiv \Upsilon \circ \mathcal{R} \circ \Upsilon^\dagger~,
\ee
the chord Yang–Baxter relation can be recast in the compact form:
\be
\begin{split}
    \mathcal{R} \circ q^{\hat{N}^{(2)}_{\text{tot}}}
    &= \Upsilon^\dagger \circ \tilde{R} \circ \Upsilon \circ q^{\hat{N}^{(2)}_{\text{tot}}}
    = \Upsilon^\dagger \circ \tilde{R}\, q^{\hat{N}^{(1)}_{\text{tot}}} \circ \Upsilon \\
    &= \Upsilon^\dagger \circ q^{\hat{N}^{(1)}_{\text{tot}}}\, \tilde{R} \circ \Upsilon
    = q^{\hat{N}^{(2)}_{\text{tot}}} \circ \Upsilon^\dagger \circ \tilde{R} \circ \Upsilon ~,
\end{split}
\ee
where the first equality follows from the definition of $\tilde{R}$, the second and fourth from~\eqref{eq:simple-rep}, and the third from the commutativity of $\tilde{R}$ with all operators acting within a one-particle channel.

This reformulation shows that the chord Yang–Baxter relation emerges directly from the fusion rules of $\uqsu$, which, as demonstrated in the previous section, fully determine the multi-particle spectrum of the theory.

\section{The Gravitational Wavefunction and its Schwarzian Limit } \label{sec:semi-classical}

In this section, we explore the bulk interpretation of gravitational wavefunctions discussed in previous section and the study the classical limit of the quantum group structure. 

\subsection{Representation on the Quantum Disk} \label{ssec:qto1}

To illustrate the discrete series representation \eqref{eq:discrete-rep} and to show explicitly that it reduces to the familiar $\mathfrak{sl}_2$ representation in the $q \to 1$ limit, we examine a realization of the $\uqsu$ generators on a holomorphic function space over the quantum disk $\rho:\mathcal{D}^{+}_{\Del}\mapsto A^{2}(\mathbb{D}_{q^{2}},\dd\nu_{q^{2},\Del}(z))$:~\footnote{This notation follows the mathematical literature on $q$-weighted Bergman spaces~\cite{qBergman}. Here, ``holomorphic'' refers to functions admitting a Taylor expansion in the variable $z$. For $0<p<\infty$, the Bergman space $A^p(D)$ is the space of all holomorphic functions $f$ in $D$ for which the $p$-norm is finite.}

\begin{equation} \label{eq:rho-def}
\rho |\Delta;n] = \sqrt{\frac{(q^{4\Delta};q^2)_n}{(q^2;q^2)_n} } z^n~.
\end{equation}
The quantum disk \( \mathbb{D}_{q^{2}} \) may be regarded as a subset of the unit disk with a discrete radial direction, and is therefore naturally identified with \( \mathbb{R}_{+,q} \times S^{1} \), with $\mathbb{R}_{+,q}=\{q^{n}\,|\, n=0,1,2\dots \}$. The measure \( \dd \nu_{q^2,\Delta}(z) \) on \( \mathbb{D}_{q^{2}} \) is chosen so that the map \( \rho \) becomes unitary.
Explicitly, we have:~\footnote{Although the measure is often stated in the literature for $\Delta>\tfrac{1}{2}$ (with $0<q<1$), the construction in fact extends to the larger domain $\Delta>0$. In particular, the measure remains finite at $\Delta=\tfrac{1}{2}$, where the value of $[2\Delta-1]_{q^2} (q^2 r^2;q^2)_{2\Delta-1}$  is obtained  by taking the limit $\Delta\to\tfrac{1}{2}$. The validity of this broader range will become clear in the subsequent discussion.
}
\be
\dd\nu_{q^{2},\Del}(z)=\frac{[2\Del-1]_{q^{2}}}{2\pi}(q^{2}r^{2};q^{2})_{2\Del-1}\dd_{q^{2}}(r^{2})\dd\te~,
\ee
where we have parameterized $z=re^{i\theta}$, and the inner product of $A^{2}(\mathbb{D}_{q^2},\dd\nu_{q^{2},\Delta}(z))$ is defined via:
\be \label{eq:disk-measure}
\int_{\mathbb{D}_{q}}\dd\nu_{q^{2},\Del}(z)\overline{f(z)}g(z)=\frac{[2\Del-1]_{q^{2}}}{2\pi}\int_{0}^{\infty}\dd_{q^{2}}(r^{2})\,(q^{2}r^{2};q^{2})_{2\Del-2}\left[\int_{0}^{2\pi}\dd\te\overline{f(re^{i\te})}g(re^{i\te})\right]~,
\ee
where the radial integral with $\dd_{q^2}(r^2)$ is defined via the Jackson $q$-integral:
\begin{equation}
\int_{0}^{\infty}\dd_{q^{2}}x\,f(x)\equiv(1-q^{2})\sum_{n=0}^{\infty}q^{2n}f(q^{2n})~.
\end{equation}
We have collected some basic facts about the $q$-integral in appendix \ref{app:review}. We now show that $\rho $ is indeed unitary. It suffices to verify this on the basis functions  $f = z^{m}$  and $g = z^{n}$, for which the inner product defined in~\eqref{eq:disk-measure} evaluates to:

\be  \label{eq:to-prove}
\bra z^{m}|z^{n}\ket=\delta_{mn}\frac{\left(q^{2};q^{2}\right)_{m}}{\left(q^{4\Del};q^{2}\right)_{m}}~.
\ee
Explicitly, we have
\be \label{eq:explicit-qintegral}
\begin{split}
    \bra z^{m}|z^{n}\ket	&=\frac{[2\Del-1]_{q^{2}}}{2\pi}\int_{0}^{\infty} \dd_{q^{2}}(r^{2})\, r^{m+n}(q^{2}r^{2};q^{2})_{2\Del-2}\int_{0}^{2\pi}\dd \theta \,e^{i(n-m)\theta}\\
	&=\delta_{mn}[2\Del-1]_{q^{2}}\int_{0}^{\infty}\dd_{q^{2}}(r^{2})\,r^{2m}(q^{2}r^{2};q^{2})_{2\Del-2}~,
\end{split}
\ee
where the last integral is identified as the $q$-Beta function:
\be
\int_{0}^{\infty}r^{2m}\dd_{q^{2}}(r^{2})\,(q^{2}r^{2};q^{2})_{2\Del-2}=\frac{\Gamma_{q^{2}}(m+1)\Gamma_{q^{2}}(2\Del-1)}{\Gamma_{q^{2}}(m+2\Del)}~.
\ee
Therefore, the coefficient in \eqref{eq:explicit-qintegral} becomes:
\be \label{eq:result-qintegral}
\frac{\Gamma_{q^{2}}(m+1)\Gamma_{q^{2}}(2\Del)}{\Gamma_{q^{2}}(m+2\Del)}=\frac{\left(q^{2m+4\Del},q^{2};q^{2}\right)_{\infty}}{\left(q^{4\Del},q^{2m+2};q^{2}\right)_{\infty}}=\frac{\left(q^{2};q^{2}\right)_{m}}{\left(q^{4\Del};q^{2}\right)_{m}}~.
\ee
The result \eqref{eq:result-qintegral} agrees with \eqref{eq:to-prove}, thus confirming that $\rho$ is a unitary. 

We now analyze \eqref{eq:discrete-rep} by mapping both the generators and the states using \( \rho \).  
Given a holomorphic function \( f(z) \), we introduce the \( q \)-dilation and \( q \)-derivative operators defined by
\begin{equation}
T_{q^2}f(z) = f(q^2 z)~, \quad D_{q^2} = \frac{1 - T_{q^2}}{(1 - q^2) z}~.
\end{equation}
Under $\rho$, the representation \eqref{eq:discrete-rep} becomes
\begin{equation} \label{eq:q-lattice-rep}
\begin{aligned}
K &= q^{\Delta} T_q~, \quad K^{-1} = q^{-\Delta} T_{q^{-1}}~, \\
E &= q^{\frac{1}{2} - \Delta} \left(z^2 D_{q^2} T_{q^{-1}} + \frac{1 - q^{4\Delta}}{1 - q^2} z T_q\right)~, \\
F &= -q^{\frac{3}{2} - \Delta} D_{q^2} T_{q^{-1}}~.
\end{aligned}
\end{equation}
Taking the classical limit $q \to 1$, and set $K = q^H, q=e^{-\lambda}$, then expand \eqref{eq:q-lattice-rep} to leading non-trivial order in $\lambda$. This yields the standard differential operator representation:
\begin{equation}
H = z \partial_z + \Delta~, \quad 
E = z^2 \partial_z + 2\Delta z~, \quad 
F = -\partial_z~,
\end{equation}
which satisfies the classical $\mathfrak{sl}_2$ algebra. The corresponding wavefunction $f(z)$ resembles a primary field with conformal weight $\Delta$,\footnote{It would be interesting to develop an explicit holographic dictionary following the approach of \cite{Almheiri:2024ayc}.} which furnishes the positive discrete representation of $\mathfrak{sl}_2$.

Note that the discussion above is formulated within the positive discrete series representation, whereas recent work on quantum disk has focused primarily on the principal series, which is more relevant for the pure gravity sector of the bulk Hilbert space. For readers interested in these developments, we refer to~\cite{Almheiri:2024ayc,Mertens:2025qg1}.

\subsection{Factorizing the Gravitational Wavefunction}

In addition to the action of generators on the quantum disk, we also want to understand the gravitational interpretation of the wavefunctions in previous section that serve as building block for the representation. We find the image of the eigenfunction $|\Delta;\te]_\phi$ under $\rho$ as:
\begin{equation} \label{eq:image-rho}
\rho |\Delta; \theta]_\phi 
= \sum_{n=0}^\infty \frac{Q_n(\cos \theta | q^{2\Delta} e^{\pm i \phi}; q^2)}{(q^2;q^2)_n} z^n 
= \frac{(q^{2\Delta} e^{\pm i \phi} z; q^2)_\infty}{(z e^{\pm i \theta}; q^2)_\infty}~,
\end{equation}
with the relevant wavefunction given by the Al-Salam Chihara polynomials.  In the case without matter, an interesting formula was derived in \cite{Mertens:2025qg1} that factorizes the gravitational wavefunction in terms of convolution integral of boundary wavefunctions:
\be \label{eq:factorization-0}
\frac{H_{n}\!\left(\cos\theta\,|\,q^{2}\right)}{(q^{2};q^{2})_{n}}
= \frac{e^{-in\theta}}{(1-q^{2})(q^{2};q^{2})_{\infty}^{2}}
\int_{0}^{\infty}\!\dd_{q^{2}}x\,x^{-2i\theta/\lambda-1}
(q^{2}x;q^{2})_{\infty}(q^{2+2n}/x;q^{2})_{\infty}~,
\ee
where the right-hand side can be interpreted as the matrix element of the quantum group generator $K^{-n}$ between wavefunctions
$\phi_j^{+}(x)=x^j (q^2 /x;q^2)_\infty$ and 
$\phi_j^{-}(x)=x^{-j} (q^2 x;q^2)_\infty$, with $j=-1/2+\frac{i\theta}{\log(q^2)}$. These functions furnish the principal series representation of $\uqslr$, in which the generator $K$ acts as a dilation operator on the $q$-lattice.  
In~\cite{Mertens:2025qg1}, equation \eqref{eq:factorization-0} was further interpreted as describing the gravitational wavefunction of DSSYK in terms of a two-boundary Whittaker function, constructed from the product of two principal series representations of $\uqslr$, corresponding to the two boundary particles, similar to the Schwarzian description of JT gravity. The gravitational boundary condition is implemented by requiring that the left and right boundary wavefunctions diagonalize the parabolic elements $E^{\pm}$ of $\uqslr$, respectively.

Interestingly, the $\uqslr$ that appears in the boundary description of \cite{Mertens:2025qg1} differs from the $\uqsu$ structure we have identified within the chord algebra in the present work. This distinction may originate from the two $\mathfrak{sl}_2$ algebras that appear in the nearly-AdS$_2$ gravity~\cite{Maldacena:2016upp}. In that setting, the Hilbert space of the left and right Schwarzian particles together with bulk matter possesses an overall gauge $\mathfrak{sl}_2$ symmetry. Imposing the gauge constraints on the this Hilbert space yields the physical bulk Hilbert space. A second, physical $\mathfrak{sl}_2$ algebra can then be constructed from two-sided operators acting on this physical Hilbert space, giving rise to the Lin–Maldacena–Zhao (LMZ) algebra~\cite{Lin:2019qwu}. It is plausible that, in the double-scaled SYK model at $0<q<1$, these two $\mathfrak{sl}_2$ symmetry algebras are simultaneously deformed into $\mathcal{U}_q(\mathfrak{sl}_2)$, but with different positive structure. The gauge $\mathfrak{sl}_2$ algebra deforms into $\mathcal{U}_q(\mathfrak{sl}(2,\mathbb{R}))$, while the LMZ algebra correspondingly deforms into $\mathcal{U}_q(\mathfrak{su}(1,1))$.
We emphasize, however, that for now this should be regarded speculative, since a precise formulation of DSSYK in terms of boundary Schwarzian particles coupled to matter at arbitrary value of $0<q<1$ remains to be established. Nevertheless, we provide the following novel formula that factorizes the wavefunction in the presence of matter, in the same spirit as \eqref{eq:factorization-0}:
\be \label{eq:factorizing}
\frac{Q_{n}\!\left(\cos\theta\,|q^{2\Delta}e^{\pm i\phi};q^2 \right)}{(q^2;q^2)_{n}}
= \frac{e^{-in\theta}\,(q^{2\Delta_{+}},q^{2\Delta_{-}};q^2)_{\infty}}{(1-q^2)\,(q^2;q^2)_{\infty}^{2}}
\int_{0}^{\infty}\!\dd_{q^2}x\,x^{-\frac{i\theta}{\lambda}-1}
\frac{(q^2 x;q^2)_{\infty}(q^{2n+2}/x;q^2)_{\infty}}
{(q^{2\Delta_{+}}x,q^{2\Delta_{-}+2n}/x;q^2)_{\infty}}~.
\ee
To derive \eqref{eq:factorizing}, we start from \eqref{eq:image-rho}, and extract the wave component as:
\begin{equation} \label{eq:Qn-exp1}
\frac{Q_n (\cos\theta|q^{2\Delta}e^{\pm i\phi};q^2)}{(q^2;q^2)_n} = \int_{-\pi}^{\pi} \frac{\dd \varphi}{2 \pi} e^{-i n \varphi} 
\frac{(q^{2\Delta} e^{i \varphi \pm i \phi}; q^2)_\infty}{(e^{i \varphi \pm i \theta}; q^2)_\infty}~.
\end{equation}
For later convenience, we introduce
\begin{equation}
q^{2\Delta_\pm} \equiv q^{2\Delta} e^{\pm i(\phi - \theta)}~.
\end{equation}
The integrand of \eqref{eq:Qn-exp1} can be expressed as:
\be
\frac{(q^{2\Delta} e^{i\vp \pm i\phi})_\infty}{(e^{i\vp\pm i \theta};q^2)_\infty } =\frac{(q^{2\Delta_+} e^{i\vp +i\theta};q^2 )_\infty}{(e^{i\vp + i \theta};q^2)_\infty} \times \frac{(q^{2\Delta_-} e^{i\vp -i\theta};q^2)_\infty}{(e^{i\vp -i\theta};q^2)_\infty}~.
\ee
Using a $q$-integral identity \eqref{eq:identity} derived in appendix \ref{app:review}, the first factor in \eqref{eq:Qn-exp1} admits the following representation:
\be
\frac{(q^{{2\Delta_+}} e^{i \varphi + i \theta}; q^2)_\infty}{(e^{i \varphi + i \theta}; q^2)_\infty}
= \frac{(q^{2\Delta_+}; q^2)_\infty}{(1-q^2)(q^2;q^2)_\infty} 
\int_{0}^{\infty} \dd_{q^2} x \, x^{\frac{-i \varphi - i \theta}{\lambda} - 1} 
\frac{(q^2 x; q^2)_\infty}{(q^{2\Delta_+} x; q^2)_\infty}~,
\ee
and similarly for the other factor. Plugging this into \eqref{eq:Qn-exp1}, we find that
\be
\begin{aligned} \label{eq:Qn-exp2}
& \frac{Q_n\left(\cos \theta \mid q^{2 \Delta} e^{ \pm i \phi} ; q^2\right)}{\left(q^2 ; q^2\right)_n}=\frac{\left(q^{2 \Delta_{ \pm}} ; q^2\right)}{\left(1-q^2\right)^2\left(q^2 ; q^2\right)_{\infty}^2} \int_{-\pi}^\pi \frac{\mathrm{~d} \varphi}{2 \pi} e^{-i n \varphi}  \\
& \quad \times \int_0^{\infty} \mathrm{d}_{q^2} x \mathrm{~d}_{q^2} y \, x^{\frac{-i \varphi-i \theta}{2\lambda}-1} y^{\frac{-i \varphi+i \theta}{2\lambda}-1} \frac{\left(q^2 x, q^2 y ; q^2\right)_{\infty}}{\left(q^{2 \Delta_{+}} x, q^{2 \Delta_{-}} y ; q^2\right)_{\infty}}~.
\end{aligned}
\ee
Performing the $\varphi$-integral yields a delta function:
\begin{equation}
\int_{-\pi}^{\pi} \frac{\dd \varphi}{2\pi} e^{-i n \varphi}(x y)^{-i \varphi/2\lambda}
= \delta\left(n + \frac{1}{2\lambda} \log(x y)\right)~,
\end{equation}
which sets $y = q^{2n} / x$. Plugging this in, and integrating over $y$, we conclude with \eqref{eq:factorizing}. 

Similar to~\eqref{eq:factorization-0}, the integrand takes the form of a $q$-convolution integral of two boundary wavefunctions defined on the discrete time lattice $\mathbb{R}_{q^2}$, whose positions are correlated through $x$ and $x^{-1}$. It would be intriguing to explore whether these two functions admit an interpretation as Whittaker wavefunctions belonging to certain representations of $\uqslr$ that incorporate the ``matter weights'' $\Delta_{\pm}$. We leave this as a promising direction for future investigation.

\subsection{Exploring the Schwarzian Regime}

In addition to the $q \to 1$ limit discussed in the previous section, we are also interested in the gravitational wavefunction in the Schwarzian regime, where one expects a close connection to the corresponding wavefunctions in JT gravity coupled to $\mathfrak{sl}_2$ matter~\cite{Lin:2022rbf}. This is implemented via the following scaling limit:\footnote{The factor of $2$ comes from our convention  $q^2=e^{-2\lambda}$. }
\begin{equation} \label{eq:double-scaling}
\lambda \to 0~, \quad \text{with } \quad 2\lambda n = l \quad \text{ fixed~,}
\end{equation}
where $l$ represents the geodesic length, a rescaled version of the chord number $n$.

To facilitate the later discussion including matter, we begin by reviewing the limiting wavefunction in the absence of matter.  
A state with a fixed chord number $n$ and boundary energy $E(\theta)$ has a wavefunction expressed in terms of the $n$-th $q$-Hermite polynomial, which admits the following integral representation \cite{Mertens:2025qg1}:
\begin{equation} \label{eq:integral-H}
\frac{H_n(\cos \theta \mid q^2)}{(q^2; q^2)_n} 
= \frac{1}{(q^2; q^2)_\infty (1 - q^2)} \int_{-\pi}^{\pi} \frac{\dd \varphi}{2\pi} 
e^{-i n \varphi} (1 - q^2)^{-\frac{2i \varphi}{\lambda}} 
\Gamma_{q^2} \left( -\frac{i \varphi \pm i \theta}{2\lambda} \right),
\end{equation}
where the $\pm$ sign in the $q$-gamma function means that we are multiplying all configurations, i.e. $\Gamma_{q^2}(x\pm i y) = \Gamma_{q^2}(x-iy)\Gamma_{q^2}(x+iy)$, and the $q$-gamma function is defined by:
\begin{equation}
(1 - q)^x \Gamma_q(x) \equiv (1 - q) \frac{(q;q)_\infty}{(q^x; q)_\infty}~.
\end{equation}
 In the scaling limit \eqref{eq:double-scaling}, the integrand simplifies to:
\begin{equation}
(1 - q^2)^{-2i \varphi / \lambda} \to e^{-2i s \log \lambda}~, \quad 
\Gamma_{q^2} \left( -\frac{i \varphi \pm i \theta}{2\lambda} \right) 
\to \Gamma(-is \pm ik)~,
\end{equation}
where we have introduced the change of variables $\varphi = 2\lambda s$ and $\theta = 2\lambda k$ to denote the boundary energies. This parameterization makes clear that the limit \eqref{eq:double-scaling} effectively zooms in on the edge of the energy spectrum. The combination of prefactors in \eqref{eq:integral-H} reduces to:
\begin{equation}
e^{-i n \varphi} (1 - q^2)^{-2i \varphi / \lambda} \to e^{-i s \tilde{l}}~,
\end{equation}
where the renormalized length $\tilde{l} \equiv l + 2 \log \lambda$ emerges naturally. Thus, \eqref{eq:integral-H} becomes:
\begin{equation} \label{eq:integral-0}
2 K_{2ik}(e^{-\tilde{l}/2}) = \int_{-\infty}^{\infty} \frac{\dd s}{2\pi} 
e^{-is \tilde{l}} \Gamma(-is \pm ik)~,
\end{equation}
which is the familiar JT gravity wavefunction, interpreted as the overlap between a fixed boundary energy state and a bulk state at a given renormalized geodesic distance on the Euclidean disk. Notably,~\eqref{eq:integral-0} can be recognized as the Mellin--Barnes representation of the following expression:
\be \label{eq:Mellin-BarnesK}
2K_{2ik}(2e^{-\tilde{l}/2})
= e^{ik\tilde{l}} \int_{0}^{\infty}\dd u\, 
u^{2ik-1} \exp\!\left(-u - \frac{e^{-\tilde{l}}}{u}\right)~,
\ee
which arises from the propagator of the boundary Schwarzian particle.

We now turn to the case with a matter insertion. Our goal is to obtain the corresponding one-particle wavefunction in the Schwarzian regime as the limit of the wavefunctions defined in~\eqref{eq:eigenstate-2}. The generating function \eqref{eq:image-rho} allows the wavefunction $Q_n$ to be expressed in terms of $\Gamma_q$ functions with the following integral representation:
\begin{equation} \label{eq:integral-Q}
\frac{Q_n(\cos \theta \mid q^{2\Delta} e^{\pm i \phi}; q^2)}{(q^2; q^2)_n} 
= (1 - q^2)^{-\Delta} \int_{-\pi}^{\pi} \frac{\dd \varphi}{2\pi} 
e^{-i n \varphi} \frac{\Gamma_{q^2} \left( -\frac{i \varphi \pm i \theta}{2\lambda} \right)}
{\Gamma_{q^2} \left( \Delta - \frac{i \varphi \pm i \phi}{2\lambda} \right)}~.
\end{equation}
Setting $\phi = 2\lambda p,\thinspace \theta = 2\lambda k$, the integral expression \eqref{eq:integral-Q} motivates the following scaling limit of the gravitational wavefunction in the Schwarzian regime:
\be \label{eq:Psi-def}
\Psi_{2ik,2ip}^{\Del}(e^{-l/2})\equiv\lim_{q\to1^{-}}(1-q^{2})^{\Del}\frac{Q_{l/(-\log(q^{2}))}(\cos(k\log(q^{2}))|q^{2\Del\pm2ip};q^{2})}{\left(q^{2};q^{2}\right)_{n}}~.
\ee
The identity \eqref{eq:integral-Q} thus reduces, in this limit, to the corresponding integral representation of $\Psi^{\Delta}_{2ik,2ip}$:
\begin{equation} \label{eq:integral-Gamma}
\Psi^\Delta_{2ik,2ip}(e^{-l/2})=\int_{-\infty}^{\infty}\frac{\dd s}{2\pi}e^{-isl}\frac{\Gamma(-is\pm ik)}{\Gamma(\Del+ip-is)\Gamma(\Del-ip-is)}~,
\end{equation}
Similar to \eqref{eq:Mellin-BarnesK}, the integral \eqref{eq:integral-Gamma} admits an equivalent Mellin–-Barnes type representation:
\begin{equation} \label{eq:Mellin-Barnes-1}
\Psi^\Delta_{2ik,2ip}(z)=\frac{1}{\Gamma(\Del_{+})\Gamma(\Del_{-})}\int_{z}^{1}\frac{\dd t}{t}\thinspace t^{ik}\left(\frac{z^{2}}{t}\right)^{-ik}(1-t)^{\Del_{+}-1}\left(1-\frac{z^{2}}{t}\right)^{\Del_{-}-1}~.
\end{equation}
The above integral can be evaluated and leads to the following exact formula for the wavefunction:
\begin{equation} \label{eq:Psi-exact}
\Psi^\Delta_{2ik,2ip}(e^{-l/2})=\frac{1}{\Gamma(2\Del)}\left(1-e^{-l}\right)^{2\Del-1}e^{-(ip-\Del)l}\thinspace_{2}F_{1}\left(\Del-ip\pm ik;2\Del;1-e^{l}\right)~.
\end{equation}
where the alternating sign in the argument is understood as in \eqref{eq:assignment}. We provide a detailed derivation of this in appendix \ref{app:Q-limit}. Unlike the case without matter, the length $l$ here is not renormalized by the infinite constant $2\log \lambda$, and thus ranges from $0$ to $\infty$. As $l\to \infty$, the wavefunction $\Psi^{\Delta}_{2ik, 2ip} (l)$ behaves as an incoming and outgoing plane wave with momentum $k$:
\be
\Psi_{2ik,2ip}^{\Del}(e^{-l/2})\simeq\frac{\Gamma(2ik)}{\Gamma(\Del+ik\pm ip)}e^{ikl}+\frac{\Gamma(-2ik)}{\Gamma(\Del-ik\pm ip)}e^{-ikl}+O\left(e^{-l/2}\right)~,\quad l\to\infty~,
\ee
while for $l\to 0$ it behaves as $\Psi^{\Delta}_{2ik,2ip} \simeq l^{2\Delta-1}$ .    

Notably, $\Psi^{\Delta}_{2ik, 2ip}$ is \emph{not} the wavefunction of JT gravity coupled to $\mathfrak{sl}_2$ matter. In the latter case, one would expect the wavefunction to reduce to the pure JT gravity result~\eqref{eq:integral-0} when $\Delta = 0$, which is not true for $\Psi^{\Delta}_{2ik, 2ip}$. The difference originates already at finite $q$ in the wavefunction~\eqref{eq:integral-Q}: rather than approaching the $q$-Hermite polynomial in the $\Delta \to 0$ limit, it does so in the opposite limit $\Delta \to \infty$, as seen from~\eqref{eq:integral-H}.  

This feature has also been observed in sine–dilaton gravity coupled to an end-of-the-world (EOW) brane, where the Al-Salam–Chihara polynomials in~\eqref{eq:integral-Q} arise as wavefunctions in the semi-open channel~\cite{Blommaert:2024whf,Blommaert:2025avl}. In this setup, one endpoint of the spatial slice lies at the asymptotic boundary, while the other terminates on the EOW brane. As pointed out in~\cite{Cui:2025sgy}, the brane acts as a defect, effectively introducing a twist in the Hilbert space upon quantizing the phase space associated with semi-open channel. This twist disappears only in the $\Delta \to \infty$ limit, corresponding to an infinitely heavy particle connecting the two endpoints of the brane, pinching it off and recovering the pure disk result of sine–dilaton gravity or DSSYK.  

Our definition~\eqref{eq:Psi-def} is therefore motivated purely by the mathematical identity~\eqref{eq:integral-Q}, producing a well-defined wavefunction that survives the Schwarzian limit\footnote{A direct $q \to 1$ limit of the combination $Q_n/(q;q)_n$ gives a divergent result.} but differs from the JT gravity wavefunction with matter. This raises the intriguing possibility that sine–dilaton gravity coupled to an EOW brane may admit a semiclassical limit distinct from that of JT gravity with an EOW brane. This is particularly interesting because one typically expects pure sine–dilaton gravity to reduce to pure JT gravity in the $q \to 1$ limit, suggesting a richer structure in the matter-coupled case.

\section{Discussion and Outlook} \label{sec:discussion}

In this article, we explored the quantum group structure underlying the chord description of the double-scaled SYK model. Concretely, by utilizing a unitary intertwiner that factorizes the one-particle chord Hilbert space into a doubled zero-particle Hilbert space, we identify the quantum group generators as specific combinations of ladder operators acting on each copy. This structure allows a decomposition of the one-particle Hilbert space into factors of positive discrete series representations of $\mathcal{U}_q(\mathfrak{su}(1,1))$, and it generalizes to the multi-particle Hilbert space through the coproduct structure and fusion rules of the quantum group. 

As a direct application, we show that the chord Yang--Baxter relation follows naturally from the fusion rule of $\mathcal{U}_q(\mathfrak{su}(1,1))$, providing its microscopic origin. We then study the gravitational wavefunction coupled to matter and derive a novel factorization formula in the spirit of~\cite{Mertens:2025qg1}, which raises questions regarding the boundary perspective of the structure established previously. We also analyze the corresponding wavefunction in the Schwarzian limit.

A feature of the identification~\eqref{eq:id} is that it appears to yield divergent $E$ and $F$ generators in the $q \to 1$ limit, seemingly in contradiction with the discussion in Section~\ref{ssec:qto1}. Indeed, in the $q \to 1$ limit, $q^{\hat{N}} \to 1$ and the left and right operators reduce to independent harmonic oscillators acting on the two sides. Dividing by a factor of $(q^{-1} - q)$ thus leads to a divergence in this limit. The resolution is that states in $\mathcal{D}_\Delta^{+}$ are built from entangled states of the doubled Hilbert space, where the action of the left annihilation operator $a_L$ on these states is close to that of the right creation operator $a_R^{\dagger}$. The ratio between their difference, multiplied by $q^{\pm \hat{N}}$ respectively, and $(q^{-1} - q)$ remains finite as $q \to 1$. Consequently, we recover the classical $\mathfrak{sl}_2$ representation in this limit, as examined in Section~\ref{ssec:qto1}.

However, the fact that the $q \to 1$ limit of~\eqref{eq:id} diverges does not necessarily rule out the possibility that it reduces to a physical $\mathfrak{sl}_2$ algebra at the operator level in the classical regime, as one typically needs to take a scaled limit in addition to $q \to 1$. This has already been explored in pure DSSYK~\cite{Lin:2022rbf}, where the $q \to 1$ limit of the chord Hamiltonian diverges. By zooming in near the edge of the spectrum, one finds that the leading divergence arises from the constant ground-state energy, while the next-to-leading term becomes the bulk Hamiltonian of pure JT gravity, acting on the subspace of states without matter insertions. Therefore, we expect that a careful scaling limit is required to obtain the classical analogue of~\eqref{eq:id}. Furthermore, the identification of the quantum group generators with the ladder operators is not unique, and alternative realizations, such as the Jordan--Schwinger representation~\cite{MAN_KO_1994}, may also be possible, which we leave for future investigation.

Another subtlety is that the matter weight $\Delta$ has been treated as a constant throughout the discussion. The matter weight arises from the matter operator of the original SYK model in the double-scaled limit, and it introduces an explicit $\Delta$-dependence in the total chord number operator in~\cite{Lin_2023}, as well as in the number-difference operator $\hat{N}$ considered in the present work. Since neither the commutation relations of the chord algebra nor those of the quantum group depend explicitly on $\Delta$, it may be possible that such $\Delta$-dependence can be determined intrinsically by the Casimir element, without reference to its origin in the SYK model.

Finally, we discuss several follow-up directions for future study in this section. 

\paragraph{Clarification on the Role of Quantum Groups in DSSYK}
In this work, we derived a factorization formula~\eqref{eq:factorizing} for the Al-Salam–Chihara polynomial, which appears to describe the bulk wavefunction in the presence of matter chords. These same wavefunctions also arise from quantizing sine–dilaton gravity coupled to end-of-world branes in the semi-open channel~\cite{Blommaert:2025avl,Cui:2025sgy}.  
While the analogous matter-free case has been understood in~\cite{Mertens:2025qg1}—where the wavefunction was identified with a Whittaker function subject to constraints imposed on each boundary component—the interpretation of~\eqref{eq:factorizing} in the presence of matter remains less clear.  

In particular, the boundary wavefunctions considered in~\cite{Mertens:2025qg1} furnish principal representations of $\uqslr$, which possesses a different $*$-structure from the $\uqsu$ algebra studied here.  
In JT gravity, the coexistence of two $\mathfrak{sl}_2$ symmetries and their distinct roles were clarified in~\cite{Lin:2019qwu}: the physical Hilbert space is obtained by imposing $\mathfrak{sl}_2$ gauge constraints on the extended Hilbert space of two boundary Schwarzian particles, and a physical $\mathfrak{sl}_2$ algebra is then constructed from gauge-invariant gravitational operators.  
We speculate that upon $q$-deformation, both algebras become $\uqsl$ but acquire different positive structures, giving rise to the distinct $\uqslr$ and $\uqsu$ quantum groups. Understanding the interplay between these two deformations would be highly illuminating.  

A concrete open question is to interpret the factorization formula~\eqref{eq:factorizing} directly in terms of quantum group representation theory. It would also be intriguing to clarify the role played by the quantum group in the proposed dual gravitational description of DSSYK.

\paragraph{Supersymmetric Generalizations}
We expect the quantum group structure to extend naturally to the supersymmetric version of the double-scaled SYK model.  
The gravitational wavefunction of the $\mathcal{N}=1$ DSSYK model has been derived in~\cite{Mertens:2025qg1} from the representation theory of $\text{OSp}_q(1|2,\mathbb{R})$~\cite{Kulish1989}.  
It would be interesting to investigate how the corresponding quantum group generators emerge from the super-chord algebra, as explored in~\cite{Berkooz:2020xne,Boruch:2023bte}.  
In particular, a supersymmetric generalization of the quantum $6j$-symbol would be desirable, as its Schwarzian limit is expected to yield the crossed four-point function in supersymmetric Liouville quantum mechanics~\cite{Lin:2022zxd}, for which an analytic expression remains unknown.

\paragraph{Relation to Complex Liouville String}
A Liouville description of the double-scaled SYK model without matter was  proposed in~\cite{Lin_2023} and further developed in~\cite{Blommaert:2025eps} in terms of a complex Liouville string.  
Moreover, multi-chord generalizations have been explored in~\cite{Berkooz:2024evs,Gao:2024lve}, leading to an effective formulation involving coupled Liouville fields.  
It would be interesting to uncover the symmetry structure underlying the Liouville framework and to extend the correspondence proposed in~\cite{Blommaert:2025eps} to incorporate matter degrees of freedom.  
In particular, the chord-block decomposition presented in this work may shed light on how matter interactions are characterized in the bulk description.   It would also be valuable to interpret the quantum $R$-matrix from a gravitational perspective, following the spirit of~\cite{Lam_2018}, which relates the classical $R$-matrix of $SU(1,1)$, arising as the $q\!\to\!1$ limit of $\mathcal{U}_q(\mathfrak{su}(1,1))$, to components of the Dray–'t~Hooft $S$-matrix.

\section*{Acknowledgments}
We thank Ahmed Almheiri, Sergio E. Aguilar-Gutierrez,  Andreas Blommaert, Jan Boruch, Chuanxin Cui, Xi Dong, Ping Gao, Elliott Gesteau, Luca Iliesiu, Misha Isachenkov, Henry W. Lin, Thomas G. Mertens, Alexey Milekhin, Bahman Najian, Vladimir Narovlansky, Geoffrey R. Penington, Moshe Rozali,  Adrián Sánchez-Garrido, Douglas Stanford, Elisa Tabor, Gabriele Di Ubaldo, Herman Verlinde and Zhenbin Yang for helpful discussions. J.X. gratefully acknowledges the hospitality of the Berkeley Leinweber Institute for Theoretical Physics and appreciates the valuable discussions that took place during the visit. J.v.d.H. acknowledges support from the National Science and Engineering Research Council of Canada (NSERC) and the Simons Foundation via a Simons Investigator Award. E.V. acknowledges support from the Heising-Simons Foundation “Observational Signatures of Quantum Gravity” collaboration
grants 2021-2817 and 2024-5307. The work of J.X. is supported by the U.S. Department of Energy, Office of Science, Office of High Energy Physics, under Award Number DE-SC0011702. J.X. acknowledges the support by the Graduate Division Dissertation Fellowship and the Physics Department Graduate Fellowship at UCSB.

\appendix

\section{Verification of the Commutation Relation} \label{sec:commutation}
We now verify that the following combination of chord operators satisfies the commutation relations \eqref{eq:defining} of the quantum group $\uqsu$:
\begin{align}
K & = q^{\hat{N}}, \quad K^{-1} = q^{-\hat{N}}, \\
(q^{-1} - q) E & = q^{\frac{1}{2}} \left( q^{-\hat{N}} a_L^\dagger - q^{\hat{N}} a_R \right), \\
(q^{-1} - q) F & = - q^{-\frac{1}{2}} \left( q^{-\hat{N}} a_L - q^{\hat{N}} a_R^\dagger \right),
\end{align}
where we have defined $\hat{N} = \hat{n}_L - \hat{n}_R + \Delta$, assuming that the operators on the right-hand side satisfy the chord algebra commutation relations~\eqref{eq:chord-commutation}.  

It is straightforward to check that $K E = q E K$ and $K F = q^{-1} F K$, which follows directly from the commutation relations between $q^{\hat{N}}$ and the ladder operators $a_{L/R}$, $a_{L/R}^\dagger$. The nontrivial part is to verify the commutator $[E, F]$. We find
\be \label{eq:EF-commute}
\begin{aligned}
(q^{-1} - q)^{2}[E,F] 
&= -[q^{-\hat{N}}a_{L}^{\dagger}, q^{-\hat{N}}a_{L}] 
   - [q^{\hat{N}}a_{R}, q^{\hat{N}}a_{R}^{\dagger}] \\
&\quad + [q^{-\hat{N}}a_{L}^{\dagger}, q^{\hat{N}}a_{R}^{\dagger}] 
   + [q^{\hat{N}}a_{R}, q^{-\hat{N}}a_{L}]~.
\end{aligned}
\ee
The terms in the second line vanish identically:
\be
\begin{aligned}
{\left[q^{-\hat{N}} a_L^{\dagger}, q^{\hat{N}} a_R^{\dagger}\right] } 
&= q^{-\hat{N}} a_L^{\dagger} q^{\hat{N}} a_R^{\dagger} - q^{\hat{N}} a_R^{\dagger} q^{-\hat{N}} a_L^{\dagger} \\
&= q^{-1} a_L^{\dagger} a_R^{\dagger} - q^{-1} a_R^{\dagger} a_L^{\dagger} \\
&= q^{-1}\left[a_L^{\dagger}, a_R^{\dagger}\right] = 0~,
\end{aligned}
\ee
where we have used that left and right creation operators commute. Similarly, it follows that $[q^{\hat{N}}a_{R}, q^{-\hat{N}}a_{L}] = 0$.

We now focus on the terms in the first line of~\eqref{eq:EF-commute}. Moving the ladder operators to one side of $q^{\hat{N}}$, we find:
\be
q^{-\hat{N}}a_{L}q^{-\hat{N}}a_{L}^{\dagger} = q^{-2\hat{N}}q^{-1}a_{L}a_{L}^{\dagger}~, 
\qquad 
q^{-\hat{N}}a_{L}^{\dagger}q^{-\hat{N}}a_{L} = q^{-2\hat{N}}q\,a_{L}^{\dagger}a_{L}~.
\ee
This shows that the factor of $q^{\hat{N}}$ effectively twists the commutator, deforming it into a $q^2$-commutator:
\be
[q^{-\hat{N}}a_{L}, q^{-\hat{N}}a_{L}^{\dagger}] = q^{-2\hat{N}} q^{-1}[a_{L}, a_{L}^{\dagger}]_{q^{2}}~.
\ee
Substituting these results into~\eqref{eq:EF-commute}, we obtain
\be
\begin{aligned}
\left(q^{-1} - q\right)^2(E F - F E) 
&= \left(q^{-2 \hat{N}} q^{-1}\left[a_L, a_L^{\dagger}\right]_{q^2}
    - q^{2 \hat{N}} q^{-1}\left[a_R, a_R^{\dagger}\right]_{q^2}\right) \\
&= \left(q^{-1} - q\right)\left(q^{-2 \hat{N}} - q^{2 \hat{N}}\right)~.
\end{aligned}
\ee
This confirms the following commutation relation of $\uqsu$:
\be
[E,F] = \frac{K^{2} - K^{-2}}{q - q^{-1}}~.
\ee

\section{Transforming to Drinfeld--Jimbo Representation} 
\label{app:transform}

In this section, we show the connection between the representation defined in~\eqref{eq:defining} and the Drinfeld--Jimbo representation of $\uqsl$, which is the universal enveloping algebra generated by $(E^{+}, E^{-}, K', K'^{-1})$ satisfying
\be \label{eq:DJ-rep}
K' E^{\pm} = q^{\pm 2} E^{\pm} K'~, 
\quad 
K' K'^{-1} = 1 = K'^{-1} K'~, 
\quad 
[E^{+}, E^{-}] = \frac{K' - K'^{-1}}{q - q^{-1}}~.
\ee
The correspondence between $(E, F, K, K^{-1})$ and the Drinfeld--Jimbo generators is given by
\be
E^{+} = q^{1/2} K^{-1} E~, 
\quad 
E^{-} = q^{-1/2} F K~, 
\quad 
K' = K^{2}~.
\ee

It is straightforward to see that the first two relations in~\eqref{eq:DJ-rep} follow directly from~\eqref{eq:defining}. The last relation is less trivial and can be verified as follows:
\be
\begin{split}
[E^{+}, E^{-}] 
&= [K^{-1} E, F K] 
= K^{-1} E F K - F E 
\\
&= [E, F] 
= \frac{K' - K'^{-1}}{q - q^{-1}}~,
\end{split}
\ee
where in the third equality we used the fact that $K^{-1} (E F) K = E F$.

We can also check that the $*$-structure in the two representations is equivalent.  
In the Drinfeld--Jimbo representation, it takes the form
\be
* E^{+} = - q^{-1} K^{-1} E^{-}~, 
\quad 
* E^{-} = - q^{-1} K E^{+}~, 
\quad 
* K' = K'~.
\ee
The third relation follows from $*K = K$ and the compatibility condition $*(a b) = (*a)(*b)$.  
Let us verify the first one explicitly:
\be
\begin{split}
* E^{+} 
&= * (q^{1/2} K^{-1} E) 
= q^{1/2} (*E) (*K^{-1}) 
= - q^{1/2} F K^{-1} 
\\
&= - q^{1/2} F K^{-2} K 
= - q^{-1} q^{-1/2} K^{-2} F 
= - q^{-1} K'^{-1} E^{-}~,
\end{split}
\ee
The second equality for $*E^{-}$ follows analogously.  
This confirms the equivalence between the representations~\eqref{eq:defining} and~\eqref{eq:DJ-rep}.

\section{Positive Discrete Representations from Chords} \label{app:Rep}
In this section, we establish the key properties of the states defined in~\eqref{eq:statem-def}, showing that they furnish the positive discrete series representation $\mathcal{D}^{+}_{\Delta}$. We first demonstrate their orthogonality relations and then analyze in detail how the quantum group generators, constructed from chord ladder operator, act on these states. 

\subsection{Orthogonality of States} \label{app:orthonormal}
We first show orthogonality of the states $\widetilde{|\Delta;m]}$. In particular, we verify the following equation
\be \label{eq:Nm-def}
\widetilde{[\Del;m}\widetilde{|\Delta;n]}=\mathcal{N}_{m}\delta_{mn}~,
\ee
and calculate the coefficients $\mathcal{N}_{m}$. Note that
\be
|a^{\da m}q^{2\Delta\hat{n}})=\sum_{k=0}^{\infty}q^{2\Delta k}a^{\da m}|k,k)=\sum_{k=0}^{\infty}q^{2\Delta k}\sqrt{\frac{\left(q^{2};q^{2}\right)_{k+m}}{\left(q^{2};q^{2}\right)_{k}}}|k+m,k)~,
\ee 
whose overlap with energy eigenstates is thus given by:
\be
(\te_{L},\te_{R}|a^{\da m}q^{2\Delta\hat{n}})=\sum_{k=0}^{\infty}q^{2\Delta k}\frac{H_{k+m}(\cos\theta_{L}|q^{2})H_{k}(\cos\te_{R}|q^{2})}{\left(q^{2};q^{2}\right)_{k}}~.
\ee
Therefore, by inserting a complete set of energy eigenstates we find that
\be
\begin{split}
\widetilde{[\Delta;m}|& \widetilde{\Delta;n}]	=\sqrt{\frac{\left(q^{4\Del};q^{2}\right)_{m}\left(q^{4\Del};q^{2}\right)_{n}}{\left(q^{2};q^{2}\right)_{m}\left(q^{2};q^{2}\right)_{n}}}\int_{0}^{\pi}\dd\mu(\te_{L})\dd\mu\left(\te_{R}\right) \\
	&\times\sum_{k=0}^{\infty}q^{2\Del k}\frac{H_{k+m}\left(\cos\te_{L}|q^{2}\right)H_{k}\left(\cos\te_{R}|q^{2}\right)}{\left(q^{2};q^{2}\right)_{k}}\sum_{k^{\pp}=0}^{\infty}q^{2\Del k^{\pp}}\frac{H_{k^{\pp}+n}\left(\cos\te_{L}|q^{2}\right)H_{k^{\pp}}\left(\cos\te_{R}|q^{2}\right)}{\left(q^{2};q^{2}\right)_{k^{\pp}}}~.
\end{split}
\ee
The integral over $\te_R$ now sets $k=k^\pp$, using the identity:
\be
\int\dd\mu(\te_{R})H_{k}(\cos\te_{R}|q^{2})H_{k^{\pp}}(\cos\te_{R}|q^{2})=\delta_{kk^{\pp}}\left(q^{2};q^{2}\right)_{k}~.
\ee
We thus find:
\be
\widetilde{[\Delta;m}|\widetilde{\Delta;n}]	=\sqrt{\frac{\left(q^{4\Del};q^{2}\right)_{m}\left(q^{4\Del};q^{2}\right)_{n}}{\left(q^{2};q^{2}\right)_{m}\left(q^{2};q^{2}\right)_{n}}}\int_{0}^{\pi}\dd\mu(\te_{L})\sum_{k=0}^{\infty}q^{4\Del k}\frac{H_{k+m}\left(\cos\te_{L}|q^{2}\right)H_{k+n}\left(\cos\te_{L}|q^{2}\right)}{\left(q^{2};q^{2}\right)_{k}}~.
\ee
Similarly, the integral over $\theta_L$ leads to $m=n$, so it follows that
\be
\widetilde{[\Delta;m}|\widetilde{\Delta;n}]=\delta_{mn}\frac{\left(q^{4\Del};q^{2}\right)_{m}}{\left(q^{2};q^{2}\right)_{m}}\sum_{k=0}^{\infty}q^{4\Delta k}\frac{\left(q^{2};q^{2}\right)_{m+k}}{\left(q^{2};q^{2}\right)_{k}}~,
\ee
where the final sum over $k$ converges for $0\leq q <1$ and any $\Delta\geq0$. This verifies \eqref{eq:Nm-def} and defines the coefficient $\mathcal{N}_m$. 

Next, we show that states $|\Delta;m]$ defined via
\be
|\Delta;m] = V_{\Delta} \widetilde{|\Delta;m]}~,\quad V_{\Delta}=\int^{\pi}_0 \dd\mu(
\theta_L)\dd \mu(\theta_R) \frac{1}{\gamma_{\Delta}(\te_L,\te_R)} |\te_L,\te_R)(\te_L,\te_R|~,
\ee
are orthonormal. This can be shown by observing that
\be
V_{\Del}^{\da}V_{\Del}=\int\dd\mu(\te_{L})\dd\mu(\te_{R})\frac{1}{\left(\te_{L}|q^{2\Delta\hat{n}}|\te_{R}\right)}|\te_{L},\te_{R})(\te_{L},\te_{R}|~.
\ee
As a consequence, the overlap between $|\Delta;n]$ and $|\Delta;m]$ is given by:
\be
[\Delta;m|\Delta;n]=\sqrt{\frac{\left(q^{4\Del};q^{2}\right)_{m}\left(q^{4\Del};q^{2}\right)_{n}}{\left(q^{2};q^{2}\right)_{m}\left(q^{2};q^{2}\right)_{n}}}\int\dd\mu(\te_{L})\dd\mu(\te_{R})\frac{(a^{\da m}q^{2\Del\hat{n}}|\te_{L},\te_{R})(\te_{L},\te_{R}|a^{\da n}q^{2\Delta\hat{n}})}{\left(\te_{L}|q^{2\Delta\hat{n}}|\te_{R}\right)}~.
\ee 
This can be expressed in terms of the Al-Salam Chihara polynomials through the formula
\be
(\te_{L},\te_{R}|a^{\da m}q^{2\Delta\hat{n}})=(\te_{L}|q^{2\Delta\hat{n}}|\te_{R})\frac{Q_{m}(\cos\te_{L}|q^{2\Delta}e^{\pm i\te_{R}};q^{2})}{\left(q^{4\Del};q^{2}\right)_{m}}~.
\ee 
We refer to appendix C of \cite{Xu:2024gfm} for a derivation of this formula. Plugging this back into the expression for the overlap, we find that:
\be
\begin{split}
  [\Delta;m|\Delta;n]	&=\int\dd\mu(\te_{L})\dd\mu(\te_{R})(\te_{L}|q^{2\Delta\hat{n}}|\te_{R})\frac{Q_{m}(\cos\te_{L}|q^{2\Delta}e^{\pm i\te_{R}};q^{2})Q_{n}(\cos\te_{L}|q^{2\Delta}e^{\pm i\te_{R}};q^{2})}{\sqrt{\left(q^{4\Del},q^{2};q^{2}\right)_{m}\left(q^{4\Delta},q^{2};q^{2}\right)_{n}}}\\
	&=\delta_{mn}~.
\end{split}
\ee
The second equality follows from the orthogonality relation of the Al-Salam Chihara polynomials. This concludes our proof that the states $\{|\Delta;m]\}$ are orthonormal.

\subsection{Details on the Action by Generators} \label{app:recursion}
In this section, we examine the action of ladder operators on the states \( |a^{\dagger m}q^{2\Delta\hat{n}}) \), and show how it motivates the identification \eqref{eq:id}. The action of \( a_L^\dagger \) is straightforward:
\begin{equation}
a_L^{\dagger}|a^{\dagger m}q^{2\Delta\hat{n}}) = |a^{\dagger (m+1)}q^{2\Delta\hat{n}})~.
\end{equation}
Next, consider the action of \( a_L \):
\begin{equation}
\begin{aligned}
a_L|a^{\dagger m}q^{2\Delta\hat{n}}) &= a_L \sum_{k=0}^{\infty} q^{2\Delta k} \sqrt{\frac{(q^2;q^2)_{m+k}}{(q^2;q^2)_k}} |k+m,k) \\
&= \sum_{k=0}^{\infty} q^{2\Delta k} \sqrt{\frac{(q^2;q^2)_{m+k}}{(q^2;q^2)_k}} \sqrt{1 - q^{2(k+m)}} |k+m-1,k) \\
&= \sum_{k=0}^{\infty} q^{2\Delta k} (1 - q^{2(m+k)}) \sqrt{\frac{(q^2;q^2)_{m+k-1}}{(q^2;q^2)_k}} |k+m-1,k)~,
\end{aligned}
\end{equation}
where we have used the following conventions for the one-sided creation and annihilation operators:
\begin{equation}
\begin{aligned}
a_L |m_L, m_R) &= \sqrt{1 - q^{2m_L}} |m_L - 1, m_R)~, \\
a_L^\dagger |m_L, m_R) &= \sqrt{1 - q^{2m_L + 2}} |m_L + 1, m_R)~.
\end{aligned}
\end{equation}
Similar relations hold for $a_{R}$ and $a_{R}^\dagger$. The action of \( a_R \) on the states is given by:
\begin{equation}
\begin{aligned}
a_R |a^{\dagger m}q^{2\Delta\hat{n}}) &= a_R \sum_{k=0}^\infty q^{2\Delta k} \sqrt{\frac{(q^2;q^2)_{m+k}}{(q^2;q^2)_k}} |k+m,k) \\
&= \sum_{k=1}^\infty q^{2\Delta k} \sqrt{\frac{(q^2;q^2)_{m+k}}{(q^2;q^2)_k}} \sqrt{1 - q^{2k}} |k+m, k-1)~.
\end{aligned}
\end{equation}
Shifting \( k \to k+1 \), we obtain the following relation:
\begin{equation}
a_R |a^{\dagger m}q^{2\Delta\hat{n}}) = q^{2\Delta} |a^{\dagger (m+1)}q^{2\Delta\hat{n}})~.
\end{equation}
This shows that $a_R$ acts the same as $a^{\dagger}_L$ on these particular states up to a constant coefficient $q^{2\Delta}$:
\begin{equation}
a_R |a^{\dagger m}q^{2\Delta\hat{n}}) = a_L^\dagger q^{2\Delta} |a^{\dagger m}q^{2\Delta\hat{n}})~.
\end{equation}
This is a manifestation of the entanglement in \( |a^{\dagger m}q^{2\Delta\hat{n}}) \). The action of \( a_R^\dagger \) is less trivial:
\begin{equation} \label{eq:ar-dagger}
\begin{aligned}
a_R^\dagger |a^{\dagger m}q^{2\Delta\hat{n}}) &= \sum_{k=0}^{\infty} q^{2\Delta k} \sqrt{\frac{(q^2;q^2)_{m+k}}{(q^2;q^2)_k}} \sqrt{1 - q^{2(k+1)}} |k+m,k+1) \\
&= \sum_{k=1}^{\infty} (1 - q^{2k}) q^{2\Delta(k-1)} \sqrt{\frac{(q^2;q^2)_{m+k-1}}{(q^2;q^2)_k}} |k+m-1,k) \\
&= q^{-2\Delta} \sum_{k=0}^{\infty} q^{2\Delta k} (1 - q^{2k}) \sqrt{\frac{(q^2;q^2)_{m+k-1}}{(q^2;q^2)_k}} |k+m-1,k)~.
\end{aligned}
\end{equation}
In the second line, we multiplied both the numerator and denominator by a factor $\sqrt{1-q^{2k+2}}$, and shifted $k\to k-1$. The last equation of \eqref{eq:ar-dagger} cannot be simplified further. However, we find that the following combination yields a particularly simple result:
\begin{equation}
\begin{aligned}
&(a_L - q^{2\Delta + 2m} a_R^\dagger) |a^{\dagger m} q^{2\Delta\hat{n}}) \\
&= \sum_{k=0}^\infty \left[ q^{2\Delta k}(1 - q^{2(m+k)}) - q^{2\Delta k}(q^{2m} - q^{2(m+k)}) \right] \sqrt{\frac{(q^2;q^2)_{m+k-1}}{(q^2;q^2)_k}} |k+m-1,k) \\
&= \sum_{k=0}^{\infty} q^{2\Delta k} (1 - q^{2m}) \sqrt{\frac{(q^2;q^2)_{m+k-1}}{(q^2;q^2)_k}} |k+m-1,k) \\
&= (1 - q^{2m}) |a^{\dagger (m-1)} q^{2\Delta\hat{n}})~.
\end{aligned}
\end{equation}
The action of the number operator $\hat{N}$ is given by:
\begin{equation}
\hat{N} |a^{\dagger m} q^{2\Delta\hat{n}})=(\hat{n}_L -\hat{n}_R +\Delta)|a^{\dagger m} q^{2\Delta\hat{n}})=(m+\Delta)|a^{\dagger m} q^{2\Delta\hat{n}}) ~.
\end{equation}
We therefore find that:
\begin{equation}
q^{-1/2}(q^{-\hat{N}} a_L^\dagger - q^{\hat{N}} a_R) |a^{\dagger m} q^{2\Delta\hat{n}}) = q^{-\Delta - m - 1/2} (1 - q^{2m + 4\Delta}) |a^{\dagger (m+1)} q^{2\Delta\hat{n}})~,
\end{equation}
\begin{equation}
q^{1/2}(q^{-\hat{N}} a_L - q^{\hat{N}} a_R^\dagger) |a^{\dagger m} q^{2\Delta\hat{n}}) = q^{1/2 - m - \Delta} (1 - q^{2m}) |a^{\dagger (m-1)} q^{2\Delta\hat{n}})~.
\end{equation}
This motivates the definition of the following set of states used in the main text:
\begin{equation}
\widetilde{|\Delta; m]} \equiv \sqrt{ \frac{(q^{4\Delta}; q^2)_m}{(q^2; q^2)_m} } |a^{\dagger m} q^{2\Delta\hat{n}})~.
\end{equation}
By the above calculations, it follows that:
\begin{equation} \label{eq:generator-I}
q^{-1/2}(q^{-\hat{N}} a_L^\dagger - q^{\hat{N}} a_R) \widetilde{|\Delta; m]} = q^{-\Delta - m - 1/2} \sqrt{(1 - q^{2m + 4\Delta})(1 - q^{2m + 2})} \widetilde{|\Delta; m+1]}~,
\end{equation}
\begin{equation}\label{eq:generator-II}
q^{1/2}(q^{-\hat{N}} a_L - q^{\hat{N}} a_R^\dagger) \widetilde{|\Delta; m]} = q^{1/2 - m - \Delta} \sqrt{(1 - q^{2m + 4\Delta - 2})(1 - q^{2m})} \widetilde{|\Delta; m - 1]}~,
\end{equation}
\begin{equation}\label{eq:generator-III}
q^{\hat{N}} \widetilde{|\Delta; m] }= q^{\Delta + m} \widetilde{|\Delta; m]}~.
\end{equation}
It is thus natural to make the following identification:
\begin{equation}
\begin{aligned}
K &= q^{\hat{N}}, \quad K^{-1} = q^{-\hat{N}}~, \\
(q^{-1} - q) E &= q^{-1/2} (q^{-\hat{N}} a_L^\dagger - q^{\hat{N}} a_R)~, \\
(q^{-1} - q) F &= -q^{1/2} (q^{-\hat{N}} a_L - q^{\hat{N}} a_R^\dagger)~.
\end{aligned}
\end{equation}
Denoting the subspace of $\mh_0\otimes\mh_0$ spanned by $\widetilde{|\Delta;m]}$ as $\widetilde{\mathcal{D}}$, it follows that $|\Delta;m] = V_\Delta \widetilde{|\Delta;m]}$ spans the positive definite series representation $\mathcal{D}^{+}_\Delta$. Therefore, \eqref{eq:generator-I}-- \eqref{eq:generator-III} establishes that $\tilde{\mathcal{D}}$ is linearly isomorphic to $\mathcal{D}^{+}_\Delta$, with $V_{\Delta}$ serving as the intertwining map between $\tilde{\mathcal{D}}$ and $\mathcal{D}^{+}_\Delta$.

\section{The Schwarzian Limit of Al-Salam Chihara Polynomials} \label{app:Q-limit}
In this section, we provide more details regarding the Schwarzian limit of the Al-Salam Chihara polynomials. We study the function $\Psi^{\Delta}_{2ik,2ip}(e^{-l})$ defined via the following integral:
\be \label{eq:Psi-def1}
\Psi^\Delta_{2ik,2ip}(e^{-l})=\int_{-\infty}^{\infty}\frac{\dd s}{2\pi}e^{-2isl}\frac{\Gamma(-is\pm ik)}{\Gamma(\Del+ip-is)\Gamma(\Del-ip-is)}~.
\ee
The ratio of gamma functions in the integrand can be represented in terms of beta function integrals of the form:
\be
\begin{aligned}
\frac{\Gamma(-i s+i k)}{\Gamma(\Delta+i p-i s)} & =\frac{1}{\Gamma\left(\Delta_{+}\right)} \int_0^1\mathrm{~d} t_{+}\, t_{+}^{i(-s+k)-1}\left(1-t_{+}\right)^{\Delta_{+}-1}~, \\
\frac{\Gamma(-i s-i k)}{\Gamma(\Delta-i p-i s)} & =\frac{1}{\Gamma\left(\Delta_{-}\right)} \int_0^1\mathrm{~d} t_{-}\, t_{-}^{i(-s-k)-1}\left(1-t_{-}\right)^{\Delta_{-}-1}~,
\end{aligned}
\ee
where we have introduced the short-hand notation $\Del_{\pm}\equiv \Del\pm i(p-k)$. Consequently, \eqref{eq:Psi-def1} can be written as:
\be \label{eq:Psi-def2}
\Psi^\Delta_{2ik,2ip}(e^{-l})=\int_{-\infty}^{\infty}\frac{\dd s \,e^{-2isl}}{2\pi\Gamma(\Del_{+})\Gamma(\Del_{-})}\int_{0}^{1}\frac{\dd t_{+}\dd t_{-}}{t_{+}t_{-}}\thinspace t_{+}^{i(-s+k)}t_{-}^{i(-s-k)}(1-t_{+})^{\Del_{+}-1}(1-t_{-})^{\Del_{-}-1}~.
\ee
The integral over $s$ now yields a delta function:
\be
\int_{-\infty}^{\infty}\frac{\dd s}{2\pi}\, e^{-2isl}(t_{+}t_{-})^{-is}=\delta(\log t_{+}t_{-}+2l)= t_{-}\delta(t_{-}-z^{2}/t_{+})~,\quad z=e^{-l}~.
\ee
Note that for $l\geq0$ we have $0<z\leq 1$. The integral over $t_+$ is therefore supported at $z^2\leq t_+ \leq 1$. Integrating over $t_-$, we find that
\be \label{eq:Psi-integral}
\Psi^\Delta_{2ik,2ip}(z)=\frac{1}{\Gamma(\Del_{+})\Gamma(\Del_{-})}\int_{z^2}^{1}\frac{\dd t}{t}\thinspace t^{ik}\left(\frac{z^{2}}{t}\right)^{-ik}(1-t)^{\Del_{+}-1}\left(1-\frac{z^{2}}{t}\right)^{\Del_{-}-1}~.
\ee
We now perform the following change of variable:
\be
t=z^{2}+(1-z^{2})x~,
\ee
so that 
\be
\dd t = (1-z^2) \dd x~,\quad 1-t	=(1-x)(1-z^{2})~,\quad1-\frac{z^{2}}{t}=\frac{(1-z^{2})x}{z^{2}+(1-z^{2})x}~.
\ee
The integral in \eqref{eq:Psi-integral} then takes the form:
\be
\begin{aligned}
I & =\left(1-z^2\right)^{b+c-1} \int_0^1 \mathrm{~d} x\left(z^2+\left(1-z^2\right) x\right)^{a-c}(1-x)^{b-1} x^{c-1} \\
& =\left(1-z^2\right)^{b+c-1} z^{2(a-c)} \int_0^1 \mathrm{~d} x\, x^{c-1}(1-x)^{b-1}\left(1-\left(1-z^{-2}\right) x\right)^{a-c}~,
\end{aligned}
\ee 
where the relevant parameters are given by
\be
a=2ik~,\quad b=\Del_{+}~,\quad c=\Del_{-}~.
\ee
This integral agrees with the following integral representation of the hypergeometric function ${}_2F_1$:
\be
\int_{0}^{1}\mathrm{~d} x\,x^{\beta-1}(1-x)^{\gamma-\beta-1}(1-wx)^{-\alpha}=\frac{\Gamma(\beta)\Gamma(\gamma-\beta)}{\Gamma\left(\gamma\right)}\thinspace_{2}F_{1}(\alpha,\beta;\gamma;w)~,
\ee 
with coefficients
\be
\beta=c~,\quad\alpha=c-a~,\quad\gamma=b+c~,\quad w=1-z^{-2}~.
\ee 
We thus find the following expression:
\be
\Psi^\Delta_{2ik,2ip}(z)=\frac{1}{\Gamma(2\Del)}(1-z^{2})^{2\Del-1}z^{2ip-2\Del}\thinspace_{2}F_{1}\left(\Del-ip\pm ik;2\Del;1-\frac{1}{z^{2}}\right)~.
\ee
Substituting $z=e^{-l}$, we obtain the result \eqref{eq:Psi-exact} as presented in the main text.

\section{Useful Formulas on $q$-Calculus} \label{app:review}

In this section, we have collected some basic facts about $q$-integrals.  We also prove a useful $q$-integral identity involving $q$-gamma functions that allows the factorization of Al-Salam Chihara polynomials discussed in the main text.

\subsection{Review of $q$-Jackson Integral} \label{app:Jackson}

The $q$-Jackson integral provides the natural $q$-analogue of the ordinary Riemann integral and plays an essential role in $q$-calculus. It defines a discretized integration measure compatible with $q$-difference operators and naturally appears when we discuss representations of $\uqsu$ on the $q$-lattice $\mathbb{R}_{q^2}$. We review some basic properties of it in this section. 

\paragraph{$q$-Integration}

For a function $f(x)$ defined on the interval $[0,a]$ with $0<q<1$, the Jackson integral is defined as
\begin{equation}
\int_{0}^{\infty/a}\dd_q x \, f(x)\equiv (1-q)a \sum_{n=0}^{\infty} q^{n} f(a q^{n})~.
\end{equation}
We are mostly interested in the case $a=1$, where we denote the above integral as
\begin{equation}
\int_{0}^{\infty}\dd_q x\, f(x)\equiv (1-q) \sum_{n=0}^{\infty} q^{n} f(q^{n})~,
\end{equation}
provided the series on the right-hand side converges absolutely. In the limit $q \to 1^{-}$, the Jackson integral reduces to the ordinary Riemann integral:
\begin{equation}
\lim_{q \to 1^{-}} \int_{0}^{\infty/a} \dd_q x\, f(x)  = \int_{0}^{a}\dd x\, f(x)~.
\end{equation}

The Jackson integral satisfies analogues of the standard rules of integration:
\begin{itemize}
    \item \textbf{Linearity:}
    \begin{equation}
    \int_0^{\infty/a}\dd_q x\, \big( \alpha f(x) + \beta g(x) \big)
    = \alpha \int_0^{\infty/a}\dd_q x\, f(x)
    + \beta \int_0^{\infty/a} \dd_q x\,g(x)~.
    \end{equation}
    \item \textbf{$q$-Integration by Parts:}
    If $f(x)$ and $g(x)$ are sufficiently regular, then
    \begin{equation}
    \int_0^{\infty/a}\dd_q x\, f(x) D_q g(x) 
    = f(a)g(a) - f(0)g(0)
    - \int_0^{\infty/a} \dd_q x\, g(qx) D_q f(x)~,
    \end{equation}
    where the $q$-difference operator is defined by
    \begin{equation}
    D_q f(x) \equiv \frac{f(x) - f(qx)}{(1-q)x}~.
    \end{equation}
    \item \textbf{Scaling Property:}
    Under the rescaling $x \to qx$, one has
    \begin{equation}
    \int_0^{\infty/a}\dd_q x\, f(qx) = q^{-1} \int_0^{\infty/qa}\dd_q x\, f(x)~.
    \end{equation}
\end{itemize}

\paragraph{$q$-Exponential}
The $q$-exponential function is defined as:
\begin{equation}
E_q(x) \equiv \sum_{n=0}^{\infty}(-1)^n q^{\frac{n(n-1)}{2}} \frac{x^n}{(q;q)_n} = (x;q)_\infty~,
\end{equation}
where \((x;q)_\infty=\prod_{n=0}^\infty (1-xq^n)\) is the \(q\)-Pochhammer symbol. It admits a simple Jackson integral identity:
\begin{equation} \label{eq:test-1}
\int_0^{\infty}\dd_q x\, E_q(q x) = 1-q~,
\end{equation}
which follows from
\be
\int_{0}^{\infty}\dd_{q}x\thinspace\left(qx;q\right)_{\infty}=(1-q)\sum_{k=0}^{\infty}q^{k}(q^{k+1};q)_{\infty}=(1-q)\sum_{k=0}^{\infty}q^{k}\frac{\left(q;q\right)_{\infty}}{\left(q;q\right)_{k}}=1-q~,
\ee 
where we used $(q^{k+1};q)_\infty= (q;q)_\infty/(q;q)_k$ in the second equality. Equation \eqref{eq:test-1} illustrates how $q$-integration preserves $q$-analogue versions of standard integral formulas and normalization conditions.

\paragraph{\(q\)-Gamma Function and Its \(q\)-Integral Representation}

The \(q\)-gamma function \(\Gamma_q(x)\) is the \(q\)-deformation of the ordinary gamma function. It is defined for \(0<q<1\) by the assignment
\begin{equation} \label{eq:qGamma-def}
\Gamma_q(x)
\equiv (1-q)^{1-x}\,\frac{(q;q)_\infty}{(q^x;q)_\infty}~.
\end{equation}
In the classical limit \(q\to1^{-}\), one recovers the ordinary Euler gamma function:
\begin{equation}
\lim_{q\to1^-}\Gamma_q(x)=\Gamma(x)~.
\end{equation}
The function \(\Gamma_q(x)\) admits a \(q\)-analogue of Euler’s integral representation. Using the $q$-integral, one obtains
\begin{equation} \label{eq:integral-qGamma}
\Gamma_q(x) (1-q)^x
=\int_{0}^{\infty}\dd_q t\, t^{x-1}\,(q t;q)_\infty~,
\qquad (0<q<1,\ \Re(x)>0)~.
\end{equation}
Here, \((t;q)_\infty\) plays the role of a \(q\)-deformed exponential damping factor that replaces \(e^{-t}\) in the classical Euler integral
\(\Gamma(x)=\int_0^\infty\dd t\, t^{x-1} e^{-t}\). This representation makes the following recursion property manifest:
\begin{equation}
\Gamma_q(x+1)=[x]_q\,\Gamma_q(x)~,
\qquad [x]_q\equiv \frac{1-q^x}{1-q}~.
\end{equation}

\subsection{Proof of an Identity} 
We end this section by proving the following $q$-integral identity,
\begin{equation} \label{eq:identity}
\int_{0}^{\infty} \dd_q t \, t^{x-1} \frac{(qt;q)_\infty}{(at;q)_\infty} 
= (1-q) \frac{(q;q)_\infty}{(a;q)_\infty} \frac{(a q^x;q)_\infty}{(q^x;q)_\infty}~,
\end{equation}
which reduces to \eqref{eq:integral-qGamma} when $a=0$. To prove it, we use
\begin{equation}
\frac{1}{(a t; q)_\infty} = \sum_{n=0}^\infty \frac{a^n}{(q;q)_n} t^n~,
\end{equation}
to expand the denominator of the integrand. Exchanging the order of summation and $q$-integration we obtain:
\begin{equation}
\sum_{n=0}^\infty \frac{a^n}{(q;q)_n} \int_{0}^{\infty} \dd_q t \, t^{x+n-1} (q t; q)_\infty~.
\end{equation}
Each term in the above expansion is computed using:
\begin{equation} \label{eq:identity-integral}
\begin{aligned}
\int_{0}^{\infty} \dd_q t \, t^{x+n-1} (q t; q)_\infty 
&= (1-q) \sum_{m=0}^\infty q^{m(x+n)} (q^{m+1}; q)_\infty \\
&= (1-q) (q;q)_\infty \sum_{m=0}^\infty \frac{q^{m (x+n)}}{(q;q)_m} \\
&= (1-q) \frac{(q;q)_\infty}{(q^{(x+n)};q)_\infty}~.
\end{aligned}
\end{equation}
Applying \eqref{eq:identity-integral} order by order gives:
\begin{equation}
\begin{aligned}
\sum_{n=0}^\infty \frac{a^n}{(q;q)_n} (1-q) \frac{(q;q)_\infty}{(q^{x+n};q)_\infty} 
&= (1-q) (q;q)_\infty \sum_{n=0}^\infty \frac{a^n}{(q;q)_n} \frac{(q^x;q)_n}{(q^x;q)_\infty} \\
&= (1-q) \frac{(q;q)_\infty (a q^x; q)_\infty}{(a;q)_\infty (q^x; q)_\infty}~.
\end{aligned}
\end{equation}
This proves the identity \eqref{eq:identity}. 

This $q$-integral identity is useful for deriving the factorization representation of the Al-Salam Chihara polynomial. By recalling the definition of the $q$-gamma function,
\begin{equation}
(1-q)^x \Gamma_q(x) = (1-q) \frac{(q;q)_\infty}{(q^x;q)_\infty}~,
\end{equation}
we find that, setting $a = q^{\Delta}$ in \eqref{eq:identity}, leads to the following expression:
\begin{equation}
\frac{\Gamma_q(x)}{\Gamma_q(\Delta + x)} 
= \frac{(1-q)^{\Delta - 1} (q^\Delta; q)_\infty}{(q;q)_\infty} 
\int_{0}^{\infty} \dd_q t \, t^{x-1} \frac{(q t; q)_\infty}{(q^\Delta t; q)_\infty}~,
\end{equation}
which is used in the main text.


\bibliography{ref}
\bibliographystyle{JHEP}

\end{document}